\documentclass[11pt,a4paper]{article}

\usepackage{amsfonts}
\usepackage{amsmath}
\usepackage{amsbsy}
\usepackage{theorem}
\usepackage{graphicx}
\usepackage{float}
\newtheorem{definition}{Definition}

\begin{document}
\sf

\title{\bfseries  Bayesian  surface regression versus  spatial  spectral nonparametric  curve regression}

\date{}

\author{M.D. Ruiz--Medina and D. Miranda\\ IMAG - Unidad de excelencia María de Maeztu - CEX2020-001105-M}

 \maketitle
\begin{abstract}
COVID--19 incidence is analyzed at
 the provinces of some  Spanish
Communities during the period  February--October, 2020. Two infinite--dimensional regression approaches are tested.  The first one is  implemented in the regression  framework introduced in Ruiz--Medina, Miranda and Espejo \cite{RuizMedinaMirandaEspejo19}.  Specifically, a bayesian framework is adopted in the  estimation of the pure point spectrum of the temporal autocorrelation operator, characterizing the second--order structure of a surface sequence.  The second approach is formulated in the context of spatial curve regression. A nonparametric estimator of the spectral density operator,  based on
the spatial periodogram operator, is computed to approximate the spatial correlation between curves.  Dimension reduction is achieved by projection onto  the empirical eigenvectors of the  long--run spatial covariance operator. Cross--validation  procedures are  implemented to test the performance of the two functional regression approaches.
\end{abstract}

\medskip
\noindent \emph{Keywords}.  Bayesian estimation; nonparametric estimation;  spatial curve regression;  spatial periodogram operator;
spatial  spectral density operator;  surface regression.
\section{Introduction}
\label{intro} The functional linear model has been extensively studied in the Functional Data Analysis (FDA) literature (see, e.g., H\"{o}rmann and Kokoszka \cite{HormannKokoszka10};
 Horv\'ath and Kokoszka \cite{HorvathKokoszka12}; Ramsay  and Silverman \cite{RamsaySilverman05}).
Several approaches contribute to the functional linear and least--squares regression context, involving scalar/functional response, and functional regressors. Just to mention a few, we refer to  smoothing spline regression,  functional principal
component regression, or functional partial least-squares regression
  (see, e.g., Cai and  Hall \cite{CaiHall06}; Crambes, Kneip and  Sarda \cite{Crambes09};
 Cuevas \cite{Cuevas14};
 Cuevas,   Febrero and   Fraiman \cite{Cuevasetal02};
 Febrero--Bande,   Galeano  and Gonzalez-Manteiga  \cite{FebreroBande15};  Marx and  Eilers
\cite{MarxEilers99};   Ruiz-Medina \cite{RuizMedina16},
  among others).
 Morris \cite{Morris15} presents an extensive review on functional regression,  focusing on the most common techniques supported by regularization methods. Wang, Chiou and M\"uller \cite{Wangetal16}  describe  the usual  FDA methodologies, including  mean and covariance analysis,  dimension reduction techniques, like Functional Principal Component Analysis,  and  recent advances in clustering/classification,  nonlinear regression, and warping techniques for functional data.  Finally, we refer to the contribution by Jadhav, Koul and Lu \cite{Jadhavetal17}, in the multivariate functional regression context, where the effect of functional covariates on the response variable is analyzed.

One can not forget the flexible  semiparametric and nonparametric functional regression approaches (see, e.g., Ferraty and Vieu
\cite{FerratyVieu06}).
A semi-functional partial linear approach
for regression, based on nonparametric time series, is considered in
Aneiros-P\'erez  and Vieu \cite{Aneiros06};  \cite{Aneiros08}.  Particularly, kernel functional regression has been widely applied, including the case where both,  the  response and the regressors are functions   (see, e.g., Ferraty,   Keilegom  and Vieu \cite{FerratyVieu12}, and   Ferraty and Vieu  \cite{Ferraty11}). In the nonparametric framework, in the case of scalar response and functional regressors,    Ferraty, Goia, Salinelli and Vieu \cite{Ferratyetal13} present  a novel approach,  where the   choice of the optimal direction,  based on the  quadratic loss function, for projection of the regressors, and  the  link function  is achieved. A more flexible framework to model possible structural changes is contemplated in
 Goia and Vieu \cite{GoiaVieu15}, reflecting the different interaction patterns between the functional regressor and response depending on the time interval considered. The benefits of sharing  high--dimensional and functional data analysis techniques  are reflected in the special issue edited by Goia and Vieu \cite{GoiaVieu16} (see  also Gao,  Shang  and Yang \cite{Gaoetal19}).

The state--space linear framework has covered a wide range of contributions in the literature on functional time series (see Bosq \cite{Bosq00}).  Indeed, since the pioneering works
by   Cardot \cite{Cardot98};  Labbas and Mourid \cite{LabbasMourid02}; Marion and  Pumo  \cite{MarionPumo04} and  Mas \cite{Mas99}, one can find different regularized   time series predictors, accompanied by the corresponding asymptotic analysis. Several  extensions, like conditional formulations (CARH(1) models), double stochastic versions,  Banach--valued  versions, and sparse--data based applications, have been addressed in a vast literature (see   Aue, Horv\'ath and Pellatt \cite{Aueetal17}; Aue and Klepsch \cite{AueKlepsch17};
Cugliari \cite{Cugliari13}; Damon and Guillas
\cite{DamonGuillas02}; \cite{DamonGuillas05};  Didericksen and  Kokoszka \cite{Didericksen12};    El Hajj \cite{ElHajj}; Ferraty,  Van Keilegom  and Vieu \cite{FerratyVieu12};
Guillas \cite{Guillas01};
\cite{Guillas02};  H\"ormann, Horv\'ath and Reeder \cite{Hormannetal13}; Horv\'ath, Huskov\'a   and  Kokoszka
\cite{Huskova10};   Horv\'ath, Kokoszka and Rice \cite{Horvathetal14}; Kara-Terki and Mourid \cite{Kara16}; Kargin and Onatski \cite{KarginOnatski08};
 Klepsch, Klüppelberg and Wei \cite{Klepschetal17};
Kokoszka  and Reimherr  \cite{KokoszkaReimherr13}; \cite{KokoszkaReimherr13b}; Kowal, Matteson  and  Ruppert \cite{KowalMattesonRuppert13b};
Laukaitis \cite{Laukaitis08};
Liu,   Xiao and  Chen \cite{Liutal16};   Mas \cite{Mas00}; \cite{Mas02}; \cite{Mas04}; \cite{Mas07};
  Mas and Menneteau \cite{MasMenneteau03},  and Mas and Pumo \cite{MasPumo07}).

  A more general treatment,  beyond structural assumptions,  can be found in the book by Hormann  and Kokoszka \cite{HormannKokoszka12}  (see also Aue, Norinho and H\"ormann \cite{Aueetal15};  G\'orecki, H\"ormann, Horv\'ath and Kokoszka \cite{Goreckietal18};
H\"ormann, Kokoszka and Nisol  \cite{Hormannetal18}; Horv\'ath, Huskov\'a and  Rice  \cite{Horvathetal13};
   Kokoszka  and  Reimherr \cite{KokoszkaReimherr13}).  The nonparametric  functional time series framework offers interesting alternatives (see, e.g.,   Aneiros-P\'erez,  Cao  and Vilar-Fern\'andez \cite{Aneiros11};   Ezzahrioui  and  Ould--Sa{\"i}d  \cite{Ezzahrioui};   Ferraty,  Goia and  Vieu \cite{FerratyGoia02}).  Finally, we mention the  recent contributions  by Li,  Robinson and Shang \cite{Lietal20}, and  Ruiz--Medina \cite{RuizMedina19} on  long--range dependence functional time series analysis,  beyond the most extensive analyzed  weak--dependent time series scenario.

  Canale  and Ruggiero \cite{CanaleRuggiero16},  Petris  \cite{Petris18}  and Torres \emph{et al.} \cite{TorresSignes21} adopt a  bayesian framework   in the functional time series context.
The present  paper also considers  a bayesian approach in the estimation of the eigenvalues of the autocorrelation operator, characterizing the dependence structure of the error term, in the   surface regression model formulated in equation (\ref{modelreg}) below (see Ruiz-Medina,  Miranda and  Espejo \cite{RuizMedinaMirandaEspejo19}). The generalized least--squares  estimator of the infinite--dimensional regression   parameter vector is then   computed from the resulting   bayesian estimator of the inverse of the  covariance matrix operator of the error term, obeying an autoregressive hilbertian time series model.

 Functional spectral analysis  is one of the main open research areas in the current literature on functional time series. In Panaretos and Tavakoli  \cite{PanaretosTavakoli13a},
  under a weak--dependent scenario, a nonparametric estimator of the spectral density operator based on the periodogram operator is derived. The asymptotic normality of the functional discrete Fourier transform of the curve data is previously proved, under suitable functional cumulant mixing conditions,  and the summability in time of the trace norm of the elements of the  covariance operator family (see also Tavakoli  \cite{Tavakoli14}).  In Panaretos and Tavakoli \cite{PanaretosTavakoli13b}, a Karhunen--Lo\'eve--like decomposition in the temporal functional spectral domain is derived, the so--called   Cram\'er--Karhunen--Lo\'eve representation, providing a harmonic principal component analysis of functional time series (see also some recent applications in the context of functional regression in  Pham  and  Panaretos \cite{PhanPanaretos18}, and   Rubin and Panaretos \cite{RubinPanaretos20a}). In addition,  Rubin and Panaretos \cite{RubinPanaretos20b}    propose simulation techniques based on the  Cram\'er--Karhunen--Lo\'eve representation. Differences in time series dynamics are detected  by hypothesis testing in the functional spectral domain in  Tavakoli and Panaretos \cite{TavakoliPanaretos16}.
Our paper considers
 a spatial  formulation of the nonparametric estimator of the spectral density operator derived  in
Panaretos and Tavakoli  \cite{PanaretosTavakoli13a}. From this estimator,  the functional entries (kernels) of the inverse of the spatial covariance matrix operator of the  curve regression error are approximated. The resulting  plug--in  generalized  least--squares estimator of the curve regression  parameter vector is computed in the spatial functional spectral domain.

The two regression  approaches presented are tested  in a real--data example, where
  COVID--19
incidence is analyzed,  since February until October,  at the provinces of the
Spanish Communities: Andaluc\'{\i}a, Arag\'on, Asturias, Cantabria, Castilla La Mancha, Castilla--Le\'on, Catalu\~na, Comunidad de Madrid, Comunidad Valenciana, Extremadura, Galicia, la Rioja, Murcia, Navarra, Pa\'{\i}s Vasco, and Canarias.  Note that, after implementing hypothesis testing (see Bosq \cite{Bosq00}; Horv\'ath, Huskov\'a and  Rice \cite{Horvathetal13}), the last community is removed in  the  spatial curve regression analysis. The performance of both, surface and curve regressions, is tested by cross--validation.   The conclusions of our empirical study are drawn in Section \ref{conclusion}. Particularly, the observed  outperformance  of the spatial curve regression approach versus the temporal surface regression could be partially supported by the spatial weak--dependent scenario  displayed by our curve data set, and the high dimensionality inherent to the parameter space  in the bayesian functional time series framework. Furthermore, the  dimension reduction  technique implemented, based on projection onto the  eigenvectors  of the  empirical long--run spatial covariance operator, favors the computational speed. Note also that  the generalized least--squares estimator of the curve regression parameter vector is computed in the spatial spectral domain, replacing  convolutions by products of the corresponding spatial functional Fourier transforms. In the supplementary material, data visualization, and  some additional outputs of the estimation algorithms analyzed are displayed as well.



\section{Bayesian dynamical surface regression}
In the following, the  random variables introduced below are defined on  the basic probability space
 $(\Omega,\mathcal{A},P),$  and take their values  in the real separable Hilbert space $H.$
We restrict our attention to  the  dynamical functional regression model  (see Ruiz-Medina,  Miranda and  Espejo \cite{RuizMedinaMirandaEspejo19}):
\begin{equation}
Y_{n} =\mu+ X_{n}^{1}(\beta_{1}) +\dots + X_{n}^{p}(\beta_{p})+\varepsilon_{n},\quad n\in \mathbb{Z},
\label{modelreg}
\end{equation}
\noindent where $\mu \in H$ is the intercept, and   $\boldsymbol{\beta}=(\beta_{1}(\cdot),\dots \beta_{p}(\cdot))^{T}\in H^{p}$  is the functional regression parameter vector.  The operators $X_{n}^{i}\in \mathcal{S}(H),$  $i=1,\dots,p,$ are the functional regressors defining  the design matrix at each time  $n\in \mathbb{Z}.$  Here, $\mathcal{S}(H)$  denotes the space of Hilbert--Schmidt operators on $H.$ The  response $Y_{n}$ and the regression error  $\varepsilon_{n}$ lie on $H,$  for each $n\in \mathbb{Z}.$

In this paper, model (\ref{modelreg}) is interpreted   as a dynamical model for disease mapping, where the functional value of the response  $Y_{n}(\cdot )$ provides the incidence or mortality log--risk map over a spatial domain $\mathcal{D}$ at  time  $n,$ $n\in \mathbb{Z}.$ It
 is defined from a linear combination of the kernel  regressors, $X_{n}^{i}=E[(Y_{n-i}-\mu)\otimes (Y_{n-i-1}-\mu)],$  $i=1,\dots, p,$ with the functional weights $\beta_{i},$ $i=1,\dots,p,$  to be estimated, satisfying   the equation
 $$\beta_{i}(\mathbf{z})=w_{i}\mathcal{R}^{-1}_{i}(Y_{n-i-1})(\mathbf{z}),\quad  \mathbf{z}\in \mathcal{D},\ \mathcal{R}_{i}=E[(Y_{n-i-1}-\mu)\otimes (Y_{n-i-1}-\mu)],$$
\noindent for $i=1,\dots ,p,$ and  for certain unknown vector $(w_{1},\dots, w_{p})\in \mathbb{R}^{p}.$  Note that, as usual, $\otimes $ denotes the tensorial product of functions. It is well--known that for $h,g\in H,$  $h\otimes g\in \mathcal{S}(H).$

 As given in Ruiz-Medina,  Miranda and  Espejo \cite{RuizMedinaMirandaEspejo19},
for  a fixed orthonormal basis $\{\varphi_{k}\}_{k\geq 1}$ of $H,$
\begin{equation} X_{n}^{i}(\varphi_{k})(\varphi_{l})=\left\langle X_{n}^{i}(\varphi_{k}),\varphi_{l}\right\rangle_{H}=x^{i}_{k,l}(n),\quad k,l\geq 1,\ \forall n\in \mathbb{Z},\quad i=1,\dots,p.\label{regressors}\end{equation}
\noindent Indeed, since $X_{n}^{i}\in \mathcal{S}(H),$    then, $\sum_{k,l}[x^{i}_{k,l}(n)]^{2}<\infty,$ and
\begin{equation}X_{n}^{i}(f)\underset{H}{=}\sum_{k,l}x^{i}_{k,l}(n)\left\langle f,\varphi_{l}\right\rangle_{H}\varphi_{k},\quad \forall f\in H,\label{regressors2}\end{equation}
\noindent  for every  $n\in \mathbb{Z},$ $i=1,\dots,p,$ where $\underset{H}{=}$ means the equality in the norm of $H.$

We work under the assumption
\begin{equation}
E\left[\varepsilon_n |  X_{n}^{1},\dots, X_{n}^{p}\right]=0,\quad \forall n\in \mathbb{Z},
\label{erruncorrelated}
\end{equation}
\noindent on the  error term  $\varepsilon\equiv \{\varepsilon_{n},\ n\in \mathbb{Z}\},$  that here is interpreted as a weak--dependent $H$--valued  process.
Indeed,   $\varepsilon $ is assumed to be a  zero-mean Autoregressive Hilbertian process of order one  (ARH(1) process), satisfying the following state equation:
\begin{equation}\varepsilon_n=\rho(\varepsilon_{n-1})+\epsilon_{n}, \ n\in\mathbb{Z},\label{ARHerroerterm}
\end{equation}
\noindent where $\rho$ denotes the autocorrelation operator, which belongs to the space of bounded linear operators $\mathcal{L}(H)$ on $H,$ satisfying
$\|\rho\|_{\mathcal{L}(H)}^{k}<1,$ for $k\geq k_{0},$ for certain $k_{0}\in \mathbb{N}.$ We restrict our attention to the Gaussian case,  with
$\{\epsilon_{n},\ n\in \mathbb{Z}\}$ being an $H$--valued Gaussian   white noise in the strong sense.  Equivalently, $\{\epsilon_{n},\ n\in \mathbb{Z}\}$  is a sequence of independent and identically distributed $H$-valued
zero-mean Gaussian random variables with trace autocovariance operator. The underlying surface covariance structure in time of $\varepsilon $ is then characterized in terms of the autocovariance $R_{0}$ and cross--covariance $R_{1}$  operators, given by:  \begin{eqnarray}R_{0}&=&E[\varepsilon_{0}\otimes \varepsilon_{0}]=E[\varepsilon_{n}\otimes\varepsilon_{n}],\quad \forall n\in \mathbb{Z}\nonumber\\ R_{1}&=&E[\varepsilon_{0}\otimes\varepsilon_{1}]=E[\varepsilon_{n}\otimes \varepsilon_{n+1}],\quad \forall n\in \mathbb{Z}.\nonumber\end{eqnarray}

Note that, under the above model assumptions (see Ruiz-Medina,  Miranda and  Espejo \cite{RuizMedinaMirandaEspejo19}):
 \begin{eqnarray}
&& \mu_{n,\mathcal{X}}=E[Y_{n}|X_{n}^{1},\dots, X_{n}^{p}]=\mu+X_{n}^{1}(\beta_{1}) +\dots + X_{n}^{p}(\beta_{p}),\quad
 n=1,\dots,N\nonumber\\
&& \hspace*{1cm} E\left[ \varepsilon_{i}\otimes \varepsilon_{j}\right]
 = \rho^{|j-i|}R_{0},\quad i,j\in \mathbb{Z},
 \label{eqresponse}
 \end{eqnarray}
\noindent    where the last identity follows from
$$\varepsilon_{n}=\sum_{j=0}^{k}\rho^{j}\epsilon_{n-j}+\rho^{k+1}(\varepsilon_{n-k-1}),\quad k\geq 1,$$
\noindent obtained by applying  invertibility of  the ARH(1) model (see equation (3.11) in Bosq \cite{Bosq00}).

Let us consider the functional sample  $Y_{1},\dots, Y_{N}.$  The  following matrix expression characterizes the infinite--dimensional  covariance structure of the errors (see Ruiz-Medina,  Miranda and  Espejo \cite{RuizMedinaMirandaEspejo19}):
 \begin{eqnarray}\mathbf{C}&=&E\left[\left(\varepsilon_{1},\dots,  \varepsilon_{N} \right)^{T}\otimes \left(\varepsilon_{1},\dots,  \varepsilon_{N}\right)\right]\nonumber\\
  &=&\left[\begin{array}{ccccc}R_{0} & \rho R_{0} &       \rho^{2} R_{0} & \ldots & \rho^{N-1} R_{0} \\
 \rho R_{0} & R_{0} & \rho R_{0} & \ldots & \rho^{N-2} R_{0}\\
\vdots & \ldots & \ldots & \ldots & \vdots\\
\rho^{N-1} R_{0} & \rho^{N-2} R_{0}&\ldots   &\ldots & R_{0}\\
 \end{array}\right]\nonumber\\&=& \left[\begin{array}{ccccc} I & \rho  &       \rho^{2}  & \ldots & \rho^{N-1}  \\
 \rho  & I & \rho  & \ldots & \rho^{N-2} \\
\vdots & \ldots & \ldots & \ldots & \vdots\\
\rho^{N-1}  & \rho^{N-2} &\ldots   &\ldots & I\\
 \end{array}\right] \nonumber\\
 &\times &\left[\begin{array}{ccccc}R_{0} & 0 &       0& \ldots & 0\\
 0 & R_{0} & 0 & \ldots & 0\\
\vdots & \ldots & \ldots & \ldots & \vdots\\
0 & 0&\ldots   &\ldots & R_{0}\\
 \end{array}\right]= \boldsymbol{\rho}\mathbf{R}_{0}.
 \label{cocopmat}
\end{eqnarray}
\noindent Under \textbf{Assumptions  A1--A2} in Ruiz-Medina,  Miranda and  Espejo \cite{RuizMedinaMirandaEspejo19}, Lemma 1 provides the
following pure point spectral representation of the autocorrelation matrix operator $\boldsymbol{\rho}$ given in equation  (\ref{cocopmat}):
For  every $\mathbf{f}\in H^{N},$
\begin{eqnarray}
\boldsymbol{\rho}(\mathbf{f})&=&
\sum_{k\geq 1}\boldsymbol{\Psi}_{k}\left[\begin{array}{cccc}1 & \lambda_{k}(\rho) &        \ldots & \left[\lambda_{k}(\rho)\right]^{N-1}  \\
 \lambda_{k}(\rho)  &1 &  \ldots & \left[\lambda_{k}(\rho)\right]^{N-2}\\
\vdots & \ldots &  \ldots & \vdots\\
\left[\lambda_{k}(\rho)\right]^{N-1}
  & \ldots   &\ldots &1\\
 \end{array}\right]\boldsymbol{\Psi}_{k}^{\star}(\mathbf{f}),
\end{eqnarray}
\noindent where
for   $\mathbf{g}=
(g_{1},\dots,g_{N})\in H^{N},$ and $k\geq 1,$
\begin{eqnarray}&&\hspace*{-2cm} \boldsymbol{\Psi}_{k}^{\star}(\mathbf{g})=\left[\begin{array}{cccc}
\psi_{k} & 0 & \ldots & 0\\
0 &\psi_{k} & \ldots &0\\
\vdots & \ddots & \ddots &\vdots \\
0 & \ldots & \ldots &\psi_{k}\\
\end{array}\right]^{\star}\left[\begin{array}{c}g_{1}\\g_{2}\\ \vdots\\ g_{N}\end{array}\right]
\nonumber\\
&=&\left[\begin{array}{c}\left\langle g_{1},\psi_{k}\right\rangle_{H}\\
\left\langle g_{2},\psi_{k}\right\rangle_{H}\\ \vdots\\ \left\langle
g_{N},\psi_{k}\right\rangle_{H}\end{array}\right]=\left[\begin{array}{c} g_{1k}\\
g_{2k}\\ \vdots\\
g_{Nk}\end{array}\right]\nonumber\\ && \nonumber\\
&&\hspace*{-2.5cm}\boldsymbol{\Psi}_{k}\boldsymbol{\Psi}_{k}^{\star}(\mathbf{g})=
\boldsymbol{\Psi}_{k}\left[\begin{array}{c} g_{1k}\\
g_{2k}\\ \vdots\\
g_{Nk}\end{array}\right]\nonumber\\
&=&
\left[\begin{array}{cccc}
\psi_{k} & 0 & \ldots & 0\\
0 &\psi_{k} & \ldots &0\\
\vdots & \ddots & \ddots &\vdots \\
0 & \ldots & \ldots &\psi_{k}\\
\end{array}\right]\left[\begin{array}{c} g_{1k}\\
g_{2k}\\ \vdots\\
g_{Nk}\end{array}\right]
\nonumber\\
&=&
\left[\begin{array}{c} g_{1k}\psi_{k}\\
g_{2k}\psi_{k}\\ \vdots\\
  g_{Nk}\psi_{k}\end{array}\right]\nonumber\end{eqnarray}\begin{eqnarray}
  && \boldsymbol{\Psi}_{k}^{\star}\boldsymbol{\Psi}_{k}=
 \left[\begin{array}{cccc}
\left\langle \psi_{k},\psi_{k}\right\rangle_{H} & 0 & \ldots & 0\\
0 &\left\langle \psi_{k},\psi_{k}\right\rangle_{H} & \ldots &0\\
\vdots & \ddots & \ddots &\vdots \\
0 & \ldots & \ldots &\left\langle \psi_{k},\psi_{k}\right\rangle_{H}\\
\end{array}\right]
\nonumber\\
&&\hspace*{1.2cm}=\left[\begin{array}{cccc}
1 & 0 & \ldots & 0\\
0 &1 & \ldots &0\\
\vdots & \ddots & \ddots &\vdots \\
0 & \ldots & \ldots &1 \\
\end{array}\right].\nonumber\\
\label{cocopmat3}
\end{eqnarray}

Here,  $\left\{\lambda_{k}(\rho),\ k\geq 1\right\}$ and $\left\{ \psi_{k},\ k\geq 1\right\}$ denote the systems of eigenvalues and eigenvectors of the autocorrelation operator $\rho$ appearing in equation (\ref{ARHerroerterm}).
Lemma 3 in Ruiz-Medina,  Miranda and  Espejo \cite{RuizMedinaMirandaEspejo19} derives, under suitable conditions, the inverse of the covariance matrix operator  $\mathbf{C}$ in  (\ref{cocopmat}), characterizing the second--order structure of the functional  regression error term. The functional entries of this inverse operator  can be obtained from the eigenvalues and eigenvectors of the autocorrelation operator $\rho.$  Specifically,  $\mathbf{C}^{-1}$ is given  by  (see Lemma 3 in  Ruiz-Medina,  Miranda and  Espejo \cite{RuizMedinaMirandaEspejo19}):
\begin{eqnarray}
\mathbf{C}^{-1}(\mathbf{f})(\mathbf{g}) &=& \sum_{k,l}[\boldsymbol{\Psi}_{l}^{\star}(\mathbf{g})]^{T}
\mathbf{H}_{l,k}\boldsymbol{\Psi}_{k}^{\star}
(\mathbf{f})\label{eqmcoCinv}\\
\mathbf{H}_{l,k}&=&\left[\begin{array}{ccccc} a_{l,k} & b_{l,k}  & 0 & \ldots   & 0  \\
b_{l,k}  &   c_{l,k} &  b_{l,k}&  \ldots  &  0 \\
 \vdots & \ddots &  \ddots  & \ddots & \vdots\\
 0& \ldots  & b_{l,k} &   c_{l,k} & b_{l,k} \\
0 &0 &  \ldots &   b_{l,k}  &
a_{l,k}\\
 \end{array}\right],\label{sdcinvcoo}
\end{eqnarray}
\noindent where $a_{l,k},b_{l,k},c_{l,k},$ $k,l\geq 1,$  satisfy the following identities in the norm of $H:$ For every $f\in H,$
\begin{eqnarray}
\widetilde{C}_{1,1}(f)&=&\widetilde{C}_{N,N}(f)= R_{0}^{-1}(I-\rho^{2})^{-1}(f)
\nonumber\\
&=& \sum_{k,l}
 \frac{1}{1-\lambda_{k}^{2}(\rho)}R_{0}^{-1}(\psi_{k})(\psi_{l})\left\langle \psi_{k},f\right\rangle_{H}\psi_{l}\nonumber\\
 &=& \sum_{k,l}a_{l,k}\left\langle \psi_{k},f\right\rangle_{H}\psi_{l}\nonumber\\
\widetilde{C}_{i,i+1}(f)&=&\widetilde{C}_{j,j-1}(f)= -R_{0}^{-1}(I-\rho^{2})^{-1}\rho (f)
\nonumber\\
&=& -\sum_{k,l}\frac{\lambda_{k}(\rho)}{1-\lambda_{k}^{2}(\rho)}
 R_{0}^{-1}(\psi_{k})(\psi_{l})\left\langle \psi_{k},f\right\rangle_{H}\psi_{l}\nonumber\\
 &=& \sum_{k,l}b_{l,k}\left\langle \psi_{k},f\right\rangle_{H}\psi_{l} ,\quad i=1,\dots, N-1,
\ j=2,\dots, N\nonumber\\
\widetilde{C}_{i,i}(f)&=& R_{0}^{-1}(I-\rho^{2})^{-1}(I+\rho^{2})(f)
\nonumber\\
 &=&\sum_{k,l}
 \frac{1+\lambda_{k}^{2}(\rho)}{1-\lambda_{k}^{2}(\rho)}
 R_{0}^{-1}(\psi_{k})(\psi_{l})
  \left\langle \psi_{k},f\right\rangle_{H}\psi_{l}\nonumber\\
 &=& \sum_{k,l}c_{l,k}\left\langle \psi_{k},f\right\rangle_{H}\psi_{l},\quad i=2,\dots, N-1.
   \label{sefecinh}
 \end{eqnarray}
\subsection{Bayesian estimation}

As given in equation (24) in  Ruiz-Medina,  Miranda and  Espejo \cite{RuizMedinaMirandaEspejo19},  the generalized least--squares estimator  $\widehat{\boldsymbol{\beta }}_{N}$ of the parameter vector $\boldsymbol{\beta }\in H^{p},$ can be computed from equations  (\ref{sdcinvcoo})--(\ref{sefecinh}) as follows:

\begin{eqnarray} \widehat{\boldsymbol{\beta }}_{N}&:=&\min_{\boldsymbol{\beta}\in H^{p}} L^{2}(\boldsymbol{\beta})=\min_{\boldsymbol{\beta}\in H^{p}}\|\mathbf{Y}-\mathbf{X}(\boldsymbol{\beta })\|_{\mathcal{H}(\boldsymbol{\varepsilon})}^{2}\nonumber \\
&=&\min_{\boldsymbol{\beta}\in H^{p}}(\mathbf{Y}-\mathbf{X}(\boldsymbol{\beta }))^{T}\mathbf{C}^{-1}(\mathbf{Y}-\mathbf{X}(\boldsymbol{\beta }))\nonumber \\
&=&\min_{\boldsymbol{\beta}\in H^{p}}\sum_{k,l}[\boldsymbol{\Psi}_{l}^{\star}(\mathbf{Y}-\mathbf{X}(\boldsymbol{\beta })]^{T}
\mathbf{H}_{l,k}\boldsymbol{\Psi}_{k}^{\star}(\mathbf{Y}-\mathbf{X}(\boldsymbol{\beta })).
\label{lossfunction}
\end{eqnarray}

We propose here a Bayesian estimation of ·$\lambda_{k}(\rho)$ in equation (\ref{sefecinh}), for every $k\geq 1.$ Hence, the entries of matrixes $\mathbf{H}_{l,k},$ $k,l\geq 1,$ are approximated from equation (\ref{sefecinh}) by replacing $R_{0}$ by its empirical version, given by  $\widehat{R}_{0}^{(N)}=\frac{1}{N}\sum_{t=1}^{N}\varepsilon_{t}\otimes \varepsilon_{t},$ and $\lambda_{k}(\rho),$ $k\geq 1,$ by their bayesian estimates.  Indeed, a truncated version of equation (\ref{sefecinh}) is considered.  Specifically,  we consider the truncated pure point  spectral  diagonal expansion
\begin{equation}\widehat{R}_{0}^{(k(N))}=\sum_{k=1}^{k(N)}\widehat{\lambda }_{k,N}\widehat{\phi}_{k,N}\otimes \widehat{\phi}_{k,N},\label{revco}
\end{equation}
\noindent where  $\widehat{R}_{0}^{(k(N))}\widehat{\phi}_{k,N}=\widehat{\lambda }_{k,N}\widehat{\phi}_{k,N},$  for $k=1,\dots,k(N).$
Here,  $k(N)<N$ such that $k(N)/N\to 0,$ $N\to \infty,$  with a certain velocity decay to ensure strong--consistency (see Bosq
 \cite{Bosq00}). Usually, $k(N)=\ln (N)$ is a suitable choice.  For $k=1,\dots , k(N),$ the bayesian estimator $\widehat{\lambda}_{k}(\rho)$   of $\lambda_{k}(\rho)$   is computed by maximizing  the posterior probability density. Namely, for $$\Delta_{\rho}(\boldsymbol{\varepsilon})= \left\{\varepsilon_{2}(\psi_{k})-\lambda_{k}(\rho )\varepsilon_{1}(\psi_{k}),\dots,
\varepsilon_{N}(\psi_{k})-\lambda_{k}(\rho )\varepsilon_{N-1}(\psi_{k})\right\}_{k=1,\dots,k(N)},$$ \noindent  with $\boldsymbol{\varepsilon}=(\varepsilon_{1},\dots,\varepsilon_{N}),$
 and  for  $\boldsymbol{\lambda}(\rho)=(\lambda_{1}(\rho ),\dots,\lambda_{k(N)}(\rho )),$  under the Gaussian distribution of the errors,
the posterior probability density   $\widetilde{L}_{k(N)}\left(\boldsymbol{\lambda}(\rho)/\boldsymbol{\varepsilon}\right)=\widetilde{L}_{k(N)}\left(\boldsymbol{\lambda}(\rho)/
\Delta_{\rho}(\boldsymbol{\varepsilon})\right)$  can be written
 as
\begin{eqnarray}&&
\widetilde{L}_{k(N)}\left(\lambda_{1}(\rho ),\dots,\lambda_{k(N)}(\rho )/\Delta_{\rho}(\boldsymbol{\varepsilon})\right)
\nonumber\\
&&\simeq L_{N}\left(\Delta_{\rho}(\boldsymbol{\varepsilon})/\lambda_{1}(\rho ),\dots,\lambda_{k(N)}(\rho )\right)
p_{k(N)}\left(\lambda_{1}(\rho ),\dots,\lambda_{k(N)}(\rho )\right)
\nonumber\\
&&=\prod_{k=1}^{k(N)}\left[\frac{1}{\sigma_{k}^{N}  (2\pi)^{N/2}} \exp \left( -\frac{1}{2\sigma^2_{k}}\sum_{t=1}^{N}\left[\epsilon_{t}(\psi_{k})\right]^2 \right)\right.\nonumber\\
& &\hspace*{1cm}\left.\times \left[\lambda_{k}(\rho )\right]^{a_{k}-1}\left(1-\lambda_{k}(\rho )\right)^{b_{k}-1} \frac{\mathbb{I}_{\{0<\lambda_{k}(\rho )<1\}}}{\mathbb{B}(a_{k}, b_{k})}\right],
\label{eqbayest}
\end{eqnarray}
 \noindent where we  work under the assumption that $\boldsymbol{\lambda}(\rho)=(\lambda_{1}(\rho ),\dots,\lambda_{k(N)}(\rho ))$  is a vector of  $k(N)$
 independent   beta  random variables with  respective shape
    parameters $a_{k}$ and  $ b_{k},$   $k=1,\dots,k(N),$ under the joint prior probability density  $p_{k(N)},$ where,  as usual, $\mathbb{I}_{0<\cdot<1}$
    denotes the indicator  function on the interval $(0,1),$ and   $\mathbb{B}(a_{k}, b_{k})$ is the beta
    function,
    $$ \mathbb{B}(a_{k}, b_{k})=\frac{\Gamma(a_{k})\Gamma(b_{k})}{\Gamma(a_{k}+b_{k})},\quad k=1,\dots, k(N).$$
\noindent     We have also applied the independence of the components of the innovation process  $\epsilon$  under
the Gaussian  strong--white noise assumption. Hence, for  each $k=1,\dots, k(N),$
   $\varepsilon_{t}(\psi_{k})=\left\langle \varepsilon_{t},\psi_{k}\right\rangle_{H},$ and
    $\sigma_{k} = \sqrt{E[\varepsilon_t(\psi_{k})]^2},$  for $t=1,\dots,N.$
In  (\ref{eqbayest}), $\simeq $  means identity except a positive constant  $\mathcal{K},$ since $\widetilde{L}_{k(N)}\left(\lambda_{1}(\rho ),\dots,\lambda_{k(N)}(\rho )/\Delta_{\rho}(\boldsymbol{\varepsilon})\right)$ is  proportional to
the  likelihood
function  $L_{N}(\Delta_{\rho}(\boldsymbol{\varepsilon})/\boldsymbol{\lambda}(\rho))$ and  the prior joint  probability density $p_{k(N)}\left(\boldsymbol{\lambda}(\rho)\right).$
 Thus, $\mathcal{K}$ is given by
 $$\mathcal{K}=\int_{\boldsymbol{\Lambda }} L_{N}(\boldsymbol{\varepsilon}/\boldsymbol{\lambda}(\rho))p_{k(N)}\left(\boldsymbol{\lambda}(\rho)\right)
 d\boldsymbol{\lambda}(\rho).$$

For $i,j=1,\dots,N,$ the following approximation  $\widehat{\widetilde{C}}_{i,j}$ to $\widetilde{C}_{i,j}$
   in (\ref{sefecinh})  is  obtained, by replacing $R_{0}$  by its truncated  empirical version    $\widehat{R}_{0}^{(k(N))}$ in (\ref{revco}), and
   $\lambda_{k}(\rho)$ by its bayesian estimate $\widehat{\lambda}_{k}(\rho)$,  for $k=1,\dots,k(N),$
\begin{eqnarray}&&
\widehat{\widetilde{C}}_{1,1}(f)=\widehat{\widetilde{C}}_{N,N}(f)
= [\widehat{R}_{0}^{(k(N))}]^{-1}(I-\widehat{\rho}_{k(N)}^{2})^{-1}(f)
\nonumber\\
&& =\sum_{k,l=1}^{k(N)}
 \frac{1}{1-\widehat{\lambda}_{k}^{2}(\rho)}\left[\widehat{R}_{0}^{(k(N))}\right]^{-1}(\psi_{k})(\psi_{l})\left\langle \psi_{k},f\right\rangle_{H}\psi_{l}\nonumber\\
 && =\sum_{k,l=1}^{k(N)}\widehat{a_{l,k}^{(N)}}\left\langle \psi_{k},f\right\rangle_{H}\psi_{l} \nonumber\\
&&\widehat{\widetilde{C}}_{i,i+1}(f)=\widehat{\widetilde{C}}_{j,j-1}(f)=
   -[\widehat{R}_{0}^{(k(N))}]^{-1}(I-\widehat{\rho}_{k(N)}^{2})^{-1}
\widehat{\rho}_{k(N)} (f)
\nonumber\\
&&= -\sum_{k,l}^{k(N)}\frac{\widehat{\lambda}_{k}(\rho)}{1-\widehat{\lambda}_{k}^{2}(\rho)}
 [\widehat{R}_{0}^{(k(N))}]^{-1}(\psi_{k})(\psi_{l})\left\langle \psi_{k},f\right\rangle_{H}\psi_{l}\nonumber\\
 && =\sum_{k,l}^{k(N)}\widehat{b_{l,k}^{(N)}}\left\langle \psi_{k},f\right\rangle_{H}\psi_{l},\quad i=1,\dots, N-1,
\ j=2,\dots, N\nonumber\\&&
\widehat{\widetilde{C}}_{i,i}(f)=
  [\widehat{R}_{0}^{(k(N))}]^{-1}(I-\widehat{\rho}_{k(N)}^{2})^{-1}(I+\widehat{\rho}_{k(N)}^{2})(f)
\nonumber\\
&&=\sum_{k,l=1}^{k(N)}
 \frac{1+\widehat{\lambda}_{k}^{2}(\rho)}{1-\widehat{\lambda}_{k}^{2}(\rho)}
[ \widehat{R}_{0}^{(k(N))}]^{-1}(\psi_{k})(\psi_{l})
  \left\langle \psi_{k},f\right\rangle_{H}\psi_{l}\nonumber\\
 && =\sum_{k,l}^{k(N)}\widehat{c_{l,k}^{(N)}}\left\langle \psi_{k},f\right\rangle_{H}\psi_{l} ,\quad i=2,\dots, N-1,
   \label{sefecinh2}
 \end{eqnarray}
\noindent for any $f\in H,$  where $\widehat{\rho}_{k(N)}=\sum_{k=1}^{k(N)}\widehat{\lambda}_{k}(\rho)\psi_{k}\otimes \psi_{k}.$
From equations (\ref{eqmcoCinv})--(\ref{sefecinh}), and (\ref{sefecinh2}), the inverse $\mathbf{C}^{-1}$ of the covariance matrix operator $\mathbf{C}$
is approximated by $\widehat{\mathbf{C}}^{-1}_{B, N},$  given by  \begin{equation}\widehat{\mathbf{C}}^{-1}_{B, N}=\sum_{k,l}^{k(N)}[\boldsymbol{\Psi}_{l}^{\star}(\mathbf{g})]^{T}
\widehat{H}_{l,k}^{(N)}\boldsymbol{\Psi}_{k}^{\star},\label{cbs}\end{equation} \noindent where, for $k,l=1,\dots,k(N),$  $\widehat{H}_{l,k}^{(N)}$ has   entries
$\widehat{a_{l,k}^{(N)}},$ $\widehat{b_{l,k}^{(N)}}$   and $\widehat{c_{l,k}^{(N)}}$ computed from  (\ref{sefecinh2}).

Under \textbf{Assumptions A1--A4} in Ruiz-Medina,  Miranda and  Espejo \cite{RuizMedinaMirandaEspejo19}, the plug--in  bayesian estimator we obtain for the functional parameter vector $\boldsymbol{\beta }$ is given by

\begin{eqnarray}\widehat{\boldsymbol{\beta}}_{B,N}&=&\left(\mathbf{X}^{T}\widehat{\mathbf{C}}^{-1}_{B, N}\mathbf{X}\right)^{-1}\mathbf{X}^{T}
\widehat{\mathbf{C}}^{-1}_{B, N}(\mathbf{Y}_{N})\nonumber\\
&=&\boldsymbol{\beta}+\left(\mathbf{X}^{T}\widehat{\mathbf{C}}^{-1}_{B, N}\mathbf{X}\right)^{-1}\mathbf{X}^{T}
\widehat{\mathbf{C}}^{-1}_{B, N}(\boldsymbol{\varepsilon}_{N}).\label{glse}
\end{eqnarray}
From (\ref{glse}), the corresponding bayesian functional regression  predictor is computed as
\begin{equation}\widehat{\mathbf{Y}}_{B,N}=\mathbf{X}\widehat{\boldsymbol{\beta}}_{B,N}.\label{predb}
\end{equation}
\subsection{Estimation algorithm 1}
\label{be}
We briefly summarize the  main steps we have followed in the implementation of  the functional regression estimation methodology  above--introduced, to compute
the predictor $\widehat{\mathbf{Y}}_{B,N},$  from  the real--data set analyzed in Section \ref{ea1}  on COVID--19 incidence in some  Spanish Communities.

\begin{itemize} \item[\textbf{Step 1}] Temporal interpolation and cubic B-spline smoothing is achieved over the    COVID--19 cumulative cases step  curves located at each one of the  Spanish provinces analyzed. Their derivatives and logarithmic transform are then computed. Spatial interpolation is also implemented.
\item[\textbf{Step 2}] Bayesian componentwise estimation of the functional entries of the inverse $\mathbf{C}^{-1}$ of the covariance matrix operator $\mathbf{C}$ (see  equations  (\ref{sefecinh2})--(\ref{cbs})),
     in terms of the truncated empirical autocovariance operator, and the bayesian estimates  of the eigenvalues of the autocorrelation operator.
\item[\textbf{Step 3}] Computation of the generalized least--squares estimator $\widehat{\boldsymbol{\beta}}_{B,N}=$ \linebreak $\left(\widehat{\beta}_{B,N}^{1},\dots,\widehat{\beta}^{p}_{B,N}\right)$ of the regression parameter vector $\boldsymbol{\beta}$ from equations (\ref{lossfunction})--(\ref{glse}).
    \item[\textbf{Step 4}]  Computation of the  bayesian predictor  $\widehat{\mathbf{Y}}_{B,N}$ from \textbf{Step 3}, as given   in equation (\ref{predb}).
                 \item[\textbf{Step 5}] Model fitting is evaluated in terms of cross--validation.
\end{itemize}

\section{Spatial  functional multiple regression approach in the spectral domain}

Let  $X=\{X_{\mathbf{z}},\ \mathbf{z}\in \mathbb{Z}^{d}\}$ be a spatial functional time series  with values in the real separable Hilbert space $\mathcal{H}=L^{2}\left([\mathcal{T}_{1},\mathcal{T}_{2}],\mu (dt)\right),$  $T_{i}\in (-\infty,\infty),$ $i=1,2.$ Here, $\mu (\cdot)$ is a finite positive  measure, whose support is the time interval $[\mathcal{T}_{1},\mathcal{T}_{2}].$ For every $\mathbf{z}\in \mathbb{Z}^{d},$ $P\left[ X_{\mathbf{z}}\in L^{2}\left([\mathcal{T}_{1},\mathcal{T}_{2}],\mu (dt)\right)\right]=1,$ i.e., $X_{\mathbf{z}}$ is a random element in  $L^{2}\left([\mathcal{T}_{1},\mathcal{T}_{2}],\mu (dt)\right).$

 Assume that $X$ is stationary in space and has zero mean. The  kernels $\left\{\widetilde{r}_{\mathbf{z},\mathbf{y}},\
 \mathbf{z},\mathbf{y}\in \mathbb{Z}^{d}\right\}$
\begin{equation*}
\widetilde{r}_{\mathbf{z},\mathbf{y}}(\tau,\sigma)= E\left[X_{\mathbf{z}}(\tau)X_{\mathbf{y}}(\sigma)\right]=  r_{\mathbf{x}}(\tau,\sigma), \quad \tau,\sigma\in [\mathcal{T}_{1},\mathcal{T}_{2}], \ \mathbf{x}=\mathbf{z}-\mathbf{y}\in \mathbb{Z}^{d}
\end{equation*}
\noindent respectively define the spatial covariance operators  $\left\{ \widetilde{\mathcal{R}}_{\mathbf{z},\mathbf{y}},\ \mathbf{z},\mathbf{y}\in \mathbb{Z}^{d}\right\}.$   Thus, for $\mathbf{y},$ $\mathbf{z}\in \mathbb{Z}^{d},$ with $\mathbf{x}=\mathbf{z}-\mathbf{y},$
\begin{eqnarray}
\widetilde{\mathcal{R}}_{\mathbf{z}-\mathbf{y}}(f)(g)&=&E\left[X_{\mathbf{z}}\otimes X_{\mathbf{y}}\right](f)(g)=\mathcal{R}_{\mathbf{x}}(f)(g)\nonumber\\
&=&E\left[\left\langle X_{\mathbf{z}},g \right\rangle_{L^{2}\left([\mathcal{T}_{1},\mathcal{T}_{2}],\mu (dt)\right)} \left\langle X_{\mathbf{y}},f \right\rangle_{L^{2}\left([\mathcal{T}_{1},\mathcal{T}_{2}],\mu (dt)\right)}\right]\nonumber\\ && \forall f,g \in L^{2}\left([\mathcal{T}_{1},\mathcal{T}_{2}],\mu (dt)\right).\label{oce}\end{eqnarray}

In particular, if in equation (\ref{oce}) we consider $\mathbf{z}=\mathbf{y}$ we obtain the definition of the spatial autocovariance operator $\mathcal{R}_{\mathbf{0}}$ satisfying
$$\mathcal{R}_{\mathbf{0}}=E\left[X_{\mathbf{z}}\otimes X_{\mathbf{z}}\right]\in \mathcal{L}^{1}\left(L^{2}\left([\mathcal{T}_{1},\mathcal{T}_{2}],\mu (dt)\right)\right),\quad \forall \mathbf{z}\in \mathbb{Z}^{d},$$
\noindent  where $\mathcal{L}^{1}\left(L^{2}\left([\mathcal{T}_{1},\mathcal{T}_{2}],\mu (dt)\right)\right)$ denotes the space of trace operators on
 \linebreak $L^{2}\left([\mathcal{T}_{1},\mathcal{T}_{2}],\mu (dt)\right).$
Equivalently, $$\|\mathcal{R}_{\mathbf{0}}\|_{\mathcal{L}^{1}\left(L^{2}\left([\mathcal{T}_{1},\mathcal{T}_{2}],\mu (dt)\right)\right)}= \sum_{k\geq 1}\lambda_{k}\left(\mathcal{R}_{\mathbf{0}}\right)=E\left\|X_{\mathbf{z}}\right\|_{L^{2}\left([\mathcal{T}_{1},\mathcal{T}_{2}],\mu (dt)\right)}^{2}=\sigma_{X}^{2}<\infty,$$
\noindent with $\mathcal{R}_{\mathbf{0}}\phi_{k}=\lambda_{k}(\mathcal{R}_{\mathbf{0}})\phi_{k},$ in $L^{2}\left([\mathcal{T}_{1},\mathcal{T}_{2}],\mu (dt)\right),$  for every $k\geq 1.$ Here,   $\{\phi_{k}\}_{k\geq 1}$ and $\{\lambda_{k}(\mathcal{R}_{\mathbf{0}})\}_{k\geq 1}$ respectively denote the orthonormal system of eigenvectors and associated system of eigenvalues of operator  $\mathcal{R}_{\mathbf{0}}.$

The estimation methodology proposed is implemented in the spatial functional spectral domain. The spatial functional spectrum of $X$  is defined in terms of the spectral density operator family $\left\{\mathcal{F}_{\boldsymbol{\omega}},\ \boldsymbol{\omega}\in [-\pi,\pi]^{d}\right\},$  characterizing its spatial  second--order structure. Particularly, we consider  a family of spatial frequency varying   integral operators, whose kernels lie  in the space  $L^{2}\left([\mathcal{T}_{1},\mathcal{T}_{2}]^{2},\mu \otimes \mu(dt,ds),\mathbb{C}\right),$ and are  given by, for each $\boldsymbol{\omega }\in [-\pi,\pi]^{d},$ and $\tau,\sigma \in [\mathcal{T}_{1},\mathcal{T}_{2}],$
\begin{equation}\label{Eq1}
f_{\boldsymbol{\omega }} (\tau,\sigma) \underset{ L^{2}\left([\mathcal{T}_{1},\mathcal{T}_{2}]^{2},\mu \otimes \mu(dt,ds),\mathbb{C}\right)}{=} \frac{1}{(2\pi)^{d}} \sum_{\mathbf{x}\in\mathbb{Z}^{d}} \exp\left(-i\left\langle \boldsymbol{\omega}, \mathbf{x}\right\rangle \right)
r_{\mathbf{x}}(\tau,\sigma),
\end{equation}
\noindent where $\underset{ L^{2}\left([\mathcal{T}_{1},\mathcal{T}_{2}]^{2},\mu \otimes \mu(dt,ds),\mathbb{C}\right)}{=}$ means the identity in the norm of the space  \linebreak  $L^{2}\left([\mathcal{T}_{1},\mathcal{T}_{2}]^{2},\mu \otimes \mu(dt,ds),\mathbb{C}\right).$

For each $\boldsymbol{\omega}\in [-\pi,\pi]^{d},$
the nonparametric estimator of the spectral density operator $\mathcal{F}_{\boldsymbol{\omega}}$  we will compute later is based on the  spatial functional Discrete Fourier Transform (SfDFT),  and periodogram operator we now introduce.
\begin{definition}
The  SfDFT of $\left\{X_{\mathbf{z}}(\tau),\ \tau\in [\mathcal{T}_{1},\mathcal{T}_{2}],\ \mathbf{z}\in  [1,T]^{d}\cap  \mathbb{Z}^{d}\right\}$  is defined as
\begin{equation}
\widetilde{X}_{\boldsymbol{\omega }}^{(\mathbf{N})}(\tau) = ((2\pi)^{d} \mathbf{N})^{-1/2}\sum_{\mathbf{z}\in  [1,T]^{d}\cap  \mathbb{Z}^{d}} X_{\mathbf{z}}(\tau)\exp\left(-i\left\langle \boldsymbol{\omega}, \mathbf{z}\right\rangle \right),
\label{dftb}
\end{equation}
\noindent   for all $\tau\in [\mathcal{T}_{1}, \mathcal{T}_{2}],$ and $\boldsymbol{\omega }\in \left\{2\pi \mathbf{z}/T,\ \mathbf{z}\in [1,T-1]^{d}\right\}$  where $\mathbf{N}=T^{d},$ and the  series (\ref{dftb}) converges in the  $L^{2}([\mathcal{T}_{1},\mathcal{T}_{2}], \mu (dt), \mathbb{C})$ norm.
\end{definition}

 The periodogram operator, denoted as  $\mathcal{I}_{\boldsymbol{\omega}}^{(\mathbf{N})},$ is computed from  the SfDFT
as follows:
\begin{eqnarray}&&
\mathcal{I}^{(\mathbf{N})}_{\boldsymbol{\omega}}(\tau , \zeta)=\widetilde{X}_{\boldsymbol{\omega}}^{(\mathbf{N})}(\tau)\overline{\widetilde{X}_{\boldsymbol{\omega}}^{(\mathbf{N})}(\zeta)}
=\frac{1}{((2\pi)^{d}\mathbf{N})}\nonumber\\
\nonumber\\
&&\times \left[\sum_{\mathbf{z}\in  [1,T]^{d}\cap  \mathbb{Z}^{d}} X_{\mathbf{z}}(\tau)\exp\left(-i\left\langle \boldsymbol{\omega}, \mathbf{z}\right\rangle \right)\right]\overline{\left[\sum_{\mathbf{z}\in  [1,T]^{d}\cap  \mathbb{Z}^{d}} X_{\mathbf{z}}(\zeta)\exp\left(-i\left\langle \boldsymbol{\omega}, \mathbf{z}\right\rangle \right)\right]}\nonumber\\
&&\hspace*{3cm}
\forall (\tau ,\zeta )\in [\mathcal{T}_{1},\mathcal{T}_{2}]^{2},\ \boldsymbol{\omega} \in \left\{2\pi \mathbf{z}/T,\ \mathbf{z}\in [1,T-1]^{d}\right\},
\label{po}
\end{eqnarray}
\noindent where convergence holds  in the $L^{2}\left([\mathcal{T}_{1},\mathcal{T}_{2}]^{2}, \mu \otimes \mu (dt, ds),\mathbb{C}\right)$ norm.

We consider the following  nonparametric estimator of the spatial spectral density operator kernel:
\begin{eqnarray}
&&\widehat{f}_{\boldsymbol{\omega}}^{(\mathbf{N})}(\tau,\zeta)=\left[\frac{(2\pi)^{d}}{\mathbf{N}}\right]\sum_{\mathbf{z}\in [1,T-1]^{d}}  W^{(\mathbf{N})}\left(\boldsymbol{\omega} - \frac{2\pi \mathbf{z}}{T} \right) \mathcal{I}^{(\mathbf{N})}_{2\pi \mathbf{z}/T}(\tau,\zeta)\nonumber\\
&& \hspace*{3cm}\forall (\tau ,\zeta )\in [\mathcal{T}_{1},\mathcal{T}_{2}]^{2},\label{enp}
\end{eqnarray}

\noindent  where  the weight function $W^{(\mathbf{N})}$ is given by
\begin{equation}\label{Eq5}
W^{(\mathbf{N})}(\mathbf{z}) = \sum_{\boldsymbol{j}\in \mathbb{Z}^{d}}\frac{1}{B_{\mathbf{N}}} W\left(\frac{\mathbf{z} + 2\pi \boldsymbol{j}}{B_{\mathbf{N}}}\right),\quad \mathbf{z}\in \mathbb{R}^{d},
\end{equation}
\noindent with $B_{\mathbf{N}}$ being the positive  bandwidth parameter, and $W$ satisfying
\begin{itemize}
\item[(1)] $W$ is positive, even, and bounded in variation
\item[(2)] $W(\mathbf{x}) =0$, if  $ \|\mathbf{x}\|\geq 1$;
\item[(3)] $\int_{\mathbb{R}^{d}} \left|W(\mathbf{x})\right|^{2}d\mathbf{x} <\infty$
\item[(4)] $\int_{\mathbb{R}^{d}} W(\mathbf{x})d\mathbf{x} =1.$
\end{itemize}
Particularly, after computing the nonparametric estimator (\ref{enp}) of the spectral density operator, the functional entries of the  spatial covariance matrix operator $\mathbf{C}$ of the  curve observations  $$\mathbf{C}=\left\{\left[\begin{array}{lll}
r_{\mathbf{0}}(\tau ,\sigma ) & \dots & r_{0,\underset{d}{\dots},T-1}(\tau ,\sigma )\\
\vdots &\vdots &\vdots \\
r_{T-1,\underset{d}{\dots},0}(\tau ,\sigma )& \dots & r_{T-1,\underset{d}{\dots},T-1}(\tau ,\sigma )\\
\end{array}\right],\ (\tau,\sigma)\in [\mathcal{T}_{1},\mathcal{T}_{2}]^{2}\right\}$$

\noindent are approximated, by applying  the inverse SfDFT, obtaining
\begin{equation}\label{Eq1b}
\widehat{r}_{\mathbf{x}}(\tau,\sigma)\underset{L^{2}\left([\mathcal{T}_{1},\mathcal{T}_{2}]^{2},\mu \otimes \mu(dt,ds)\right)}{=}  \sum_{\boldsymbol{\omega}} \widehat{f}_{\boldsymbol{\omega}}^{(\mathbf{N})}(\tau,\sigma) \exp\left(i\left\langle \boldsymbol{\omega}, \mathbf{x}\right\rangle \right),
\end{equation}
\noindent for  all $\tau,\sigma \in [\mathcal{T}_{1},\mathcal{T}_{2}],$ and for each  $\mathbf{x}\in [0,T-1]^{d}.$ Thus, we obtain the estimator  $\widehat{\mathbf{C}}_{S,\mathbf{N}}$ of $\mathbf{C},$  given by
\begin{eqnarray}
&& \widehat{\mathbf{C}}_{S,\mathbf{N}}\left\{\left[\begin{array}{lll}
\widehat{r}_{\mathbf{0}}(\tau ,\sigma ) & \dots & \widehat{r}_{0,\underset{d}{\dots},T-1}(\tau ,\sigma )\\
\vdots &\vdots &\vdots \\
\widehat{r}_{T-1,\underset{d}{\dots},0}(\tau ,\sigma )& \dots & \widehat{r}_{T-1,\underset{d}{\dots},T-1}(\tau ,\sigma )\\
\end{array}\right],\ (\tau,\sigma)\in [\mathcal{T}_{1},\mathcal{T}_{2}]^{2}\right\}.\nonumber\\
\end{eqnarray}

 The plug--in  generalized least--squares estimator $\widehat{\boldsymbol{\beta} }_{S,N}$ of  $\boldsymbol{\beta },$   and  the corresponding functional regression predictor $\widehat{\mathbf{Y}}_{S,N}$  are then obtained from the following identities:

\begin{eqnarray}
\widehat{\boldsymbol{\beta} }_{S,N}&=&
\left(\mathbf{X}^{T}\widehat{\mathbf{C}}^{-1}_{S,N}\mathbf{X}\right)^{-1}\mathbf{X}^{T}
\widehat{\mathbf{C}}^{-1}_{S, N}(\mathbf{Y}_{N})\nonumber\\
\widehat{\mathbf{Y}}_{S,N}&=&\mathbf{X}\widehat{\boldsymbol{\beta} }_{S,N}
\nonumber\\ &=&
\mathbf{X}\left(\left(\mathbf{X}^{T}\widehat{\mathbf{C}}^{-1}_{S,N}\mathbf{X}\right)^{-1}\mathbf{X}^{T}
\widehat{\mathbf{C}}^{-1}_{S, N}(\mathbf{Y}_{N})\right).\nonumber\\
\label{pnsdo}\end{eqnarray}

\noindent  Here, $\mathbf{X}$ is defined from the spatial formulation of equation (\ref{modelreg}), leading to the definition of the kernel regressors $X_{\mathbf{z}}^{ij}=E[(Y_{\mathbf{z}-\mathbf{h}_{i}}-\boldsymbol{\mu})\otimes (Y_{\mathbf{z}-\mathbf{h}_{j}}-\boldsymbol{\mu})],$  with   $\mathbf{h}_{i}, \mathbf{h}_{j},$ $i,j=1,\dots, p,$
being the non--negative vectors of spatial lags, involved in the definition of the   nearest neighborhood of the curve response value at $\mathbf{z},$ keeping in mind its  significative spatial  interactions  with other spatial functional values.

\subsection{Estimation algorithm 2}
\label{ea2}
We now formulate the  main steps of the estimation algorithm implemented, to compute the spatial functional spectral predictor $\widehat{\mathbf{Y}}_{S,N},$ in the
statistical analysis  of  the COVID--19 incidence, from the  reported cases during the period February–-October, 2020,  at some  Spanish Communities.

\begin{itemize} \item[\textbf{Step 1}]  After temporal interpolation and cubic B-spline smoothing of the  cumulative cases step curves located at each one of the Spanish provinces analyzed, their derivatives and logarithmic transform are  computed. Again, spatial interpolation to a $10\times 10$ regular grid is performed.
\item[\textbf{Step 2}] Tapering the  spatiotemporal data, and compute  the empirical long--run spatial covariance operator.
\item[\textbf{Step 3}]  Compute the singular value decomposition  of the empirical  long--run spatial covariance operator obtained in \textbf{Step 2}.
\item[\textbf{Step 4}] Apply the SfDFT to the tapered spatial log--intensity   curves, after their projection onto the  selected empirical right eigenvectors of the long--run spatial covariance operator. Namely, the choice $M=5$ for the truncation parameter  is made,  explaining a 99\% of the empirical variability.
   \item[\textbf{Step 5}] Computation of the projected spatial periodogram operator is then achieved.
\item[\textbf{Step 6}]  The nonparametric estimator of the spatial spectral density operator is calculated by defining  $W$ from the modified  Bartlett--Hann window.
    \item[\textbf{Step 7}] Equation (\ref{pnsdo}) is implemented in the projected spatial functional spectral domain.
     \item[\textbf{Step 8}] The  inverse SfDFT applied to the output of \textbf{Step 7}
then leads to the spatial curve regression predictor $\widehat{\mathbf{Y}}_{S,N}$ in equation (\ref{pnsdo}).
          \item[\textbf{Step 9}] The curves at the nodes of the  first row and column of the initial $10\times 10$ regular grid are considered in the definition of the  random initial condition to run $9$--fold cross validation.
          At the $n$th iteration of this procedure ($n=1,\dots, 9$),    the curves located at the nodes in the $n$th row and $n$th column of the $9\times 9$  grid define the target spatial curve sample. The remaining curves conform the training spatial functional sample.

\end{itemize}
\section{Spanish COVID--19 incidence analysis}
Data are obtained from the declaration of COVID--19 cases by the National Epidemiological Surveillance Network (RENAVE), through the computer platform via the Web SiViES (Spanish Surveillance System), managed by the National Epidemiology Center (CNE). This information comes from the epidemiological case survey that each Autonomous Community completes upon the identification of  COVID-19 cases. The provinces and Autonomous Communities are indicated by the ISO 3166-2 code published by the International Standardization Organization (ISO).  An acceptable  quality  of the  records drives the selection procedure of the Spanish Communities analyzed  during the period  February--October, 2020.

\subsection{Estimation algorithm one}
\label{ea1}
A functional sample of size $N=1061$ of  COVID--19 incidence log--risk  surfaces, covering the area of the Spanish Communities analyzed, is obtained after applying functional data (FD) preprocessing (see \textbf{Step 1} in Section \ref{be}).  Edge effects are removed by reducing to $1000$ the number of temporal nodes defining      the surface sample size.
We find here an important difference regarding implementation of estimation algorithm 2 where data tapering is applied.

 After removing the intercept $\mu,$
least--squares $2$--D polynomial fitting  is implemented  to approximate kernel regressors in  model (\ref{modelreg}) from their empirical version, applying  'fit' MatLab function. Our polynomial choice in the argument of \emph{fit} function corresponds to the best goodness of fit reported in \emph{gof} output of \emph{fit}  function. To implement \textbf{Step 2}, based on bayesian estimation of the residual correlation structure,  ordinary least squares is first applied in terms of the computed functional design matrix, following similar steps to the ones described in Section 4 in \cite{RuizMedinaMirandaEspejo19}, under the choice $k(N)=\ln(N)$ of the truncation parameter. Note that  conditions of Proposition 1 in  \cite{RuizMedinaMirandaEspejo19} hold under the kernel polynomial fitting previously achieved.  The beta shape hyperparameters $a_{k},$ $b_{k},$  $k=1,\dots, k(N),$ for the prior in equation (\ref{eqbayest}), are selected according to the bootstrap probability density fitted  to the eigenvalues of the empirical correlation structure of the ordinary least--squares residuals (see Step 8 of the estimation algorithm proposed in Torres \emph{et al.} \cite{TorresSignes21} for the statistical analysis of COVID--19 mortality).  To compute equation (\ref{sefecinh2}), equation (\ref{eqbayest}) is maximized following a similar procedure to Step 9 in Torres \emph{et al.} \cite{TorresSignes21}, from  \emph{gaoptimset} MaLab function (selecting
\emph{HybridFcn} option). The selected option of \emph{gaoptimset}  function runs a hybrid genetic algorithm, involving quasi-Newton methodology in the optimization procedure applied after the  genetic
 algorithm finishes. The corresponding outputs allows us to implement \textbf{Step 3}, where a bayesian  approximation  (\ref{glse}) to equation     (\ref{lossfunction}), in terms of  $\widehat{\mathbf{C}}^{-1}_{B,N},$ is obtained from equations (\ref{sefecinh2})--(\ref{cbs}) computed  in \textbf{Step 2}.  \textbf{Step 4} follows straightforward from \textbf{Steps 1--3}, and equation (\ref{predb}).

 Finally,  Leave--One Out Cross Validation (LOOCV) is implemented from \textbf{Step 1--4}. Specifically, our training sample is obtained by removing one surface at each iteration of the cross--validation procedure. This surface is considered as the target output to be compared with the output of the corresponding  iteration after implementing \textbf{Steps 1--4}. Note that the reduced sample after eliminating edge effects, and removing the initial times, where the random initial conditions are defined, has size $993.$ The  $\ell^{1}$--norm of the computed   functional error at each one of the iterations is also calculated. Its mean value over the 993 iterations is reflected in  Table \ref{T1CL}, when we restrict our attention to  ten of the sixteen communities initially analyzed: Andaluc\'ia (AN) (Almer\'ia (AL), C\'adiz (CA), C\'ordoba (CO), Granada (GR), Huelva (H), Ja\'en (J), M\'alaga (MA), Sevilla (SE)); Arag\'on (AR), (Huescar (HU), Teruel (TE), Zaragoza (Z)); Castilla y Le\'on (CL) (\'Avila, (AV), Burgos (BU), Le\'on (LE), Palencia (P), Salamanca (SA), Segovia (SG), Soria (SO), Valladolid (VA), Zamora (ZA)); Castilla La Mancha (CM) (Albacete (AB), Ciudad Real (CR), Cuenca (CU), Guadalajara (GU), Toledo (TO)); Canaria (CN) (Gran Canaria (GC), Tenerife (TF));  Catalu\~na (CT) (Barcelona (B), Girona (GI), Lleida (L), Tarragona (T)); Comunidad Valenciana(VC) (Alicante (A), Castell\'on (CS), Valencia (V)); Extremadura (EX) (Badajoz (BA), C\'aceres (CC)); Galicia (GA) (A Coru\~na (C), Lugo (LU), Ourense (OR), Pontevedra (PO)); and  Pa\'is Vasco (PV) (Vizcaya (BI), Guip\'uzcoa (SS), \'Alava (VI)). One can observe  at Soria (Castilla--Le\'on)  and   Barcelona (Catalu\~na), the limit LOOCV error values  (see also Figure \ref{fig:CVALEST1} below, and  Figures 11--12 in the Supplementary Material). The   LOOCV error mean is \textbf{0.1029395349}.

 Data and \textbf{Step 4} output  visualization, in terms of  monthly averaged COVID--19 incidence maps, and their bayesian functional regression estimates, based on the overall sample,  are  displayed in Figures 1--10 in  Section 1 of the Supplementary Material.

\begin{table}[H]
\centering
\scriptsize{\begin{tabular}{c c c c c c c c c c}
Region & P1 & P2 & P3 & P4 & P5 & P6 & P7 & P8 & P9\\ \hline	\hline
AN & AL	& CA	& CO &	GR & H & J & MA & SE\\
 & 0.0321 & 0.0694 & 0.0836 & 0.1275 & 0.0217 & 0.0775 & 0.1475 & 0.2082 \\
AR & HU & TE	& Z \\
 & 0.0197 & 0.0201 & 0.1268 \\
CL & AV	& BU & LE &	P & SA & SG & SO & VA & ZA\\	
 & 0.0342 & 0.0750 & 0.0972 & 0.0370 & 0.0894 & 0.0449 & 0.0178 & 0.1505 &  0.0313\\
CM & AB & CR & CU & GU & TO\\
& 0.0511 &	0.1494 & 0.0452 & 0.0648 & 0.1939 \\
CN & GC & TF\\
& 0.0738 &	0.0553 \\
CT & B & GI & L & T\\
 & 0.9516 & 0.0843 & 0.0630  & 0.0715\\
EX & BA & CC \\
& 0.0831 &	0.0746 \\
GA & C & LU & OR & PO \\
& 0.0863 &	0.0180 & 0.0596 & 0.0617 \\
PV & BI & SS & VI \\
& 0.1938 & 0.1002 &	0.0839 \\
VC & A & CS & V \\
& 0.1383 &	0.0330 & 0.2003 \\
\end{tabular}}
\caption{\scriptsize{\emph{LOOCV  errors after running   993 iterations}}}\label{T1CL}
\end{table}
\begin{figure}[h]
\vspace{-2cm}
\begin{center}
$\begin{array}{cc}
\includegraphics[width=7.5cm,height=7.5cm]{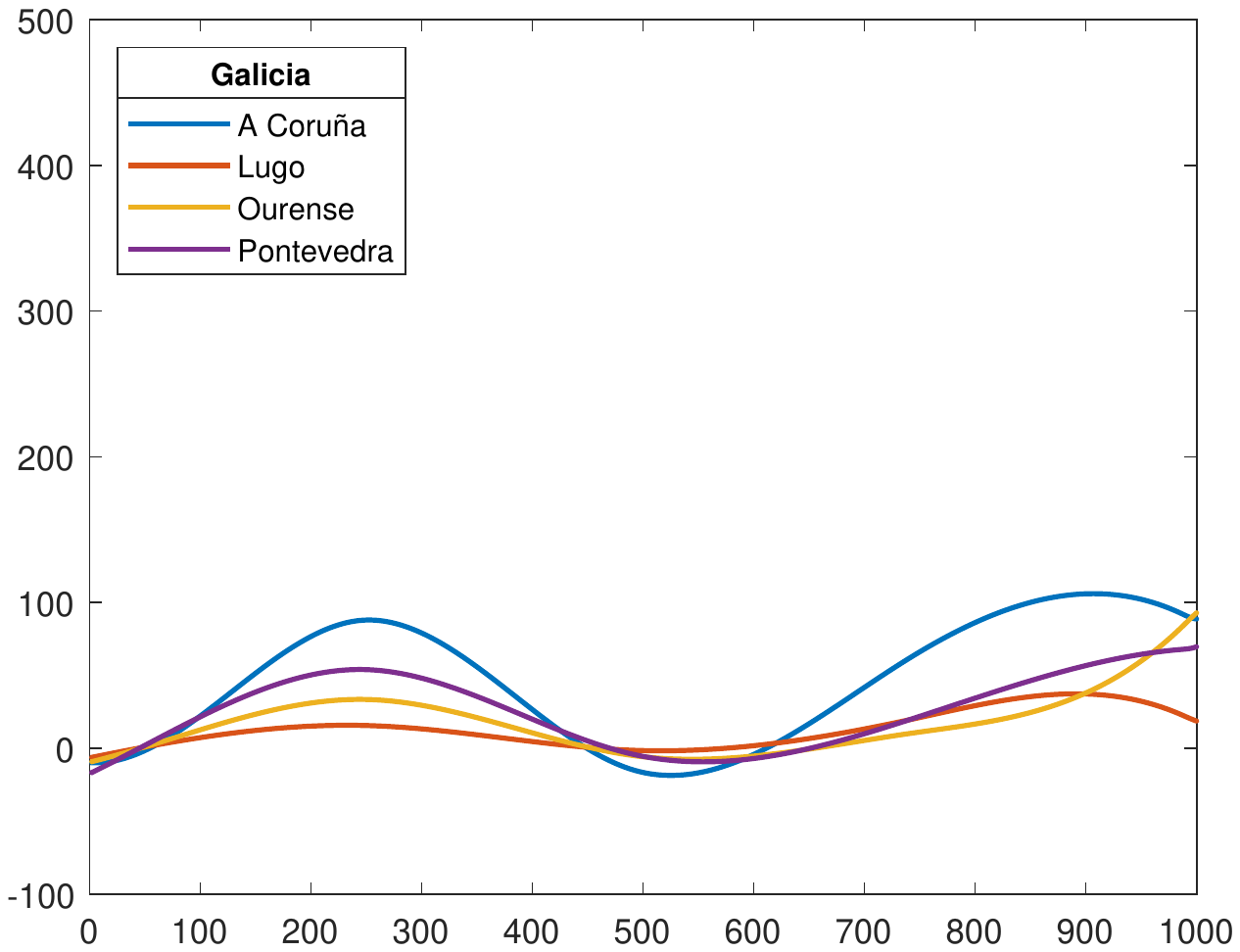}&
\hspace{-2cm}
\vspace{-4.5cm}
\includegraphics[width=7.5cm,height=7.5cm]{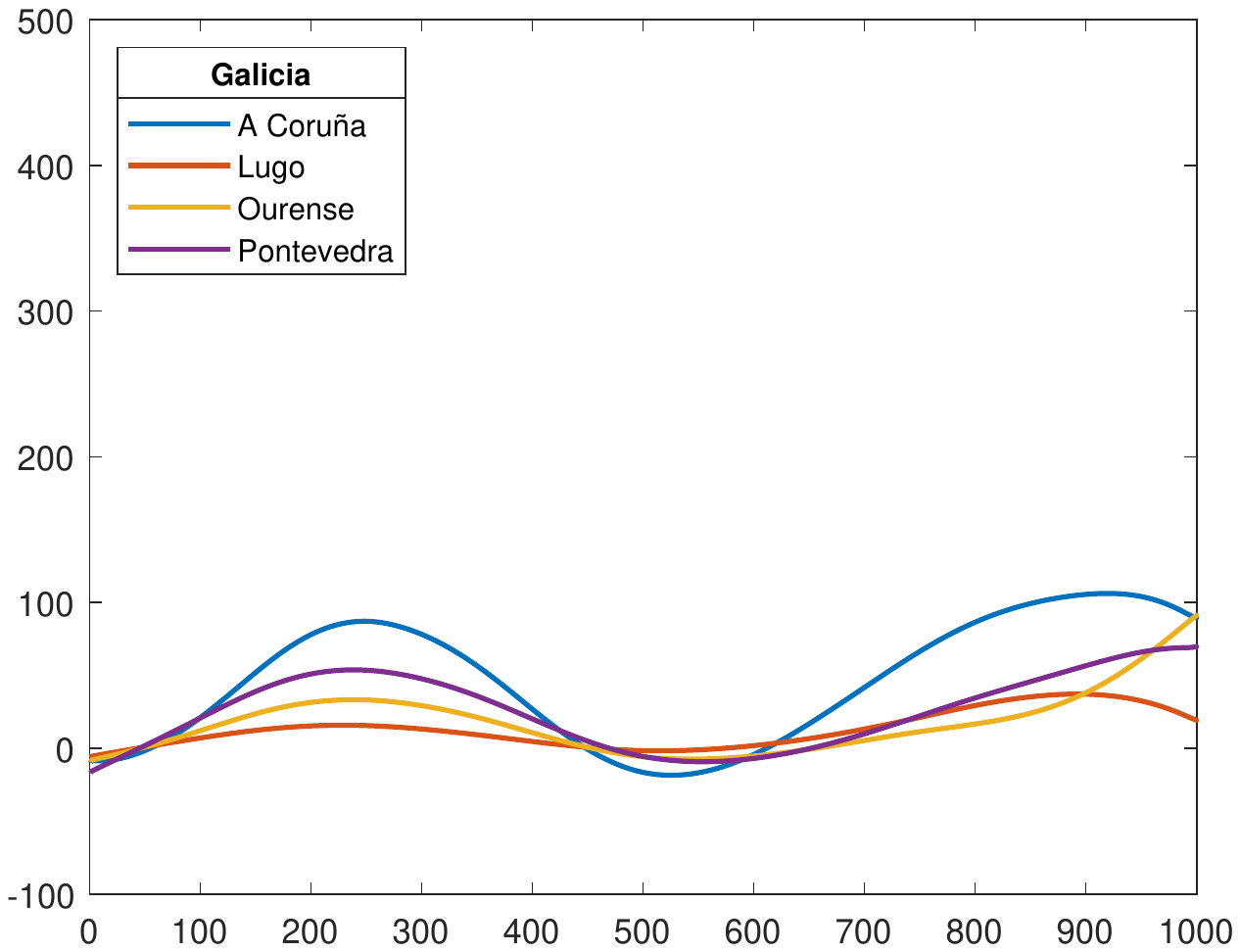}\\
\includegraphics[width=7.5cm,height=7.5cm]{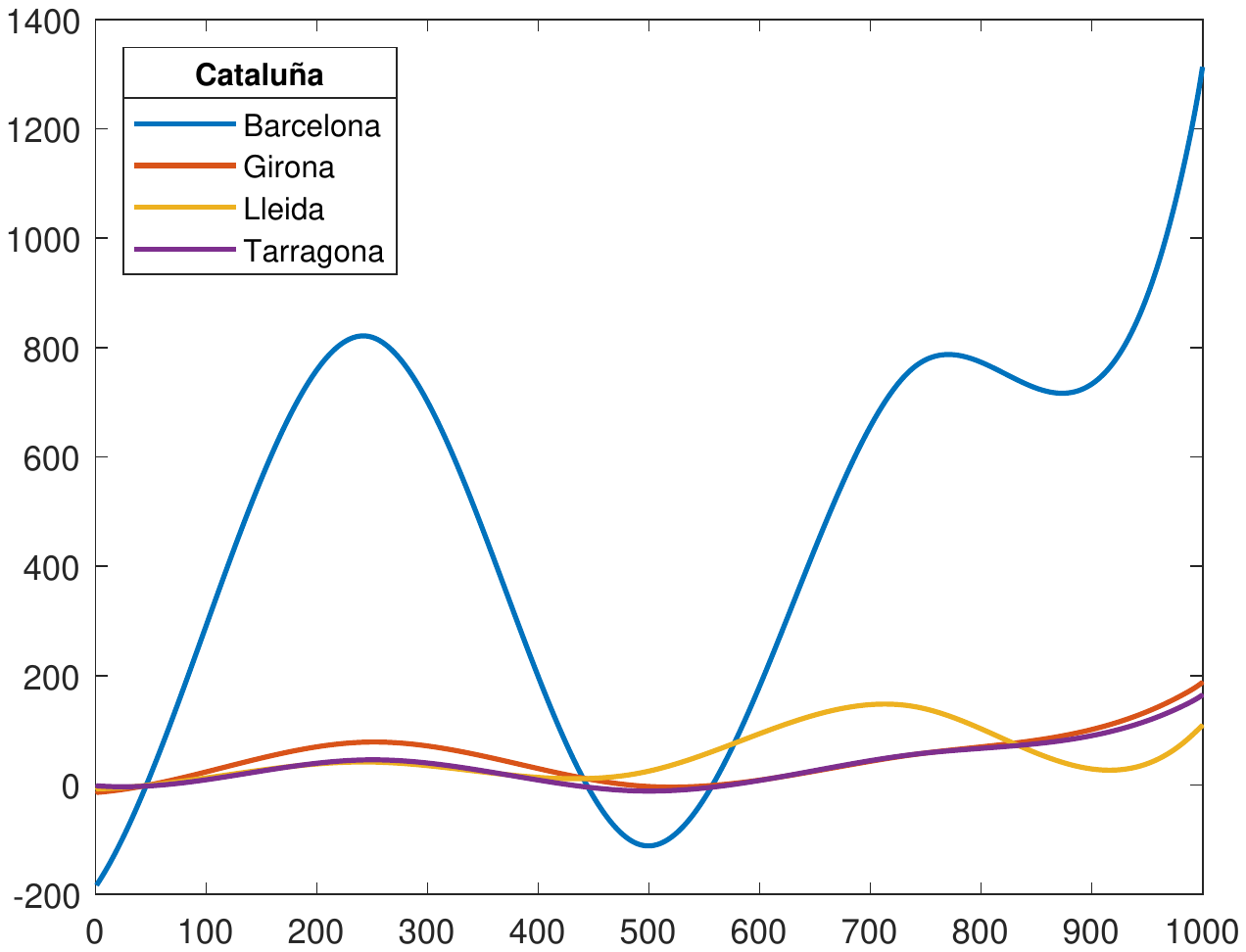}&
\hspace{-2cm}
\includegraphics[width=7.5cm,height=7.5cm]{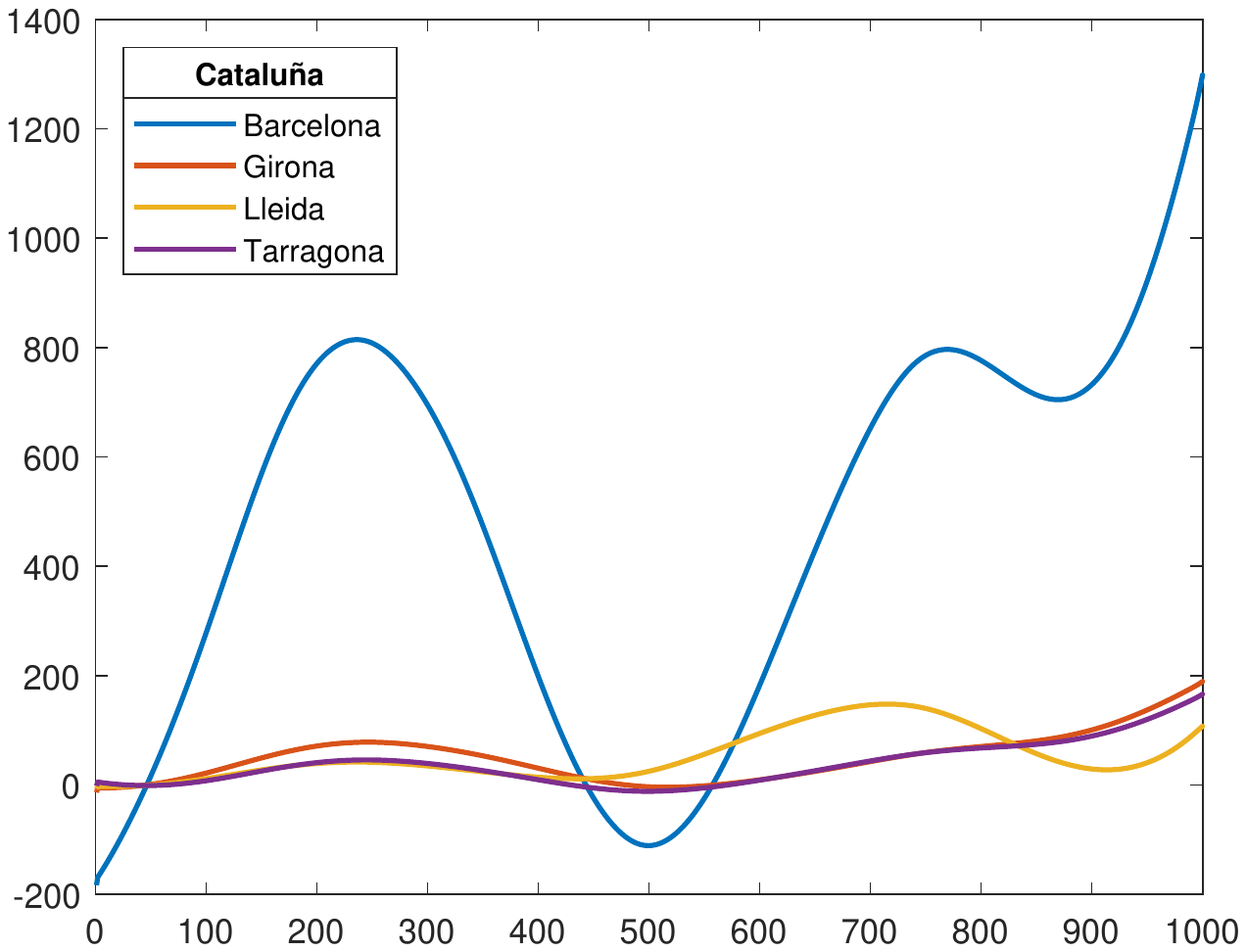}
\end{array}$
\end{center}
\vspace{-2.5cm}
\caption{\scriptsize{\emph{Original  data  (left--hand--side), and Bayesian regression predictions (right--hand--side)  at Galicia (top) and Catalu\~na (bottom)}}} \label{fig:CVALEST1}
\end{figure}

\subsection{Estimation algorithm two}
\label{esa2}
\begin{figure}[H]
\begin{center}
$\begin{array}{cc}
\vspace{-6cm}
\hspace{-1.5cm}
 \includegraphics[width=8.5cm,height=10.5cm]{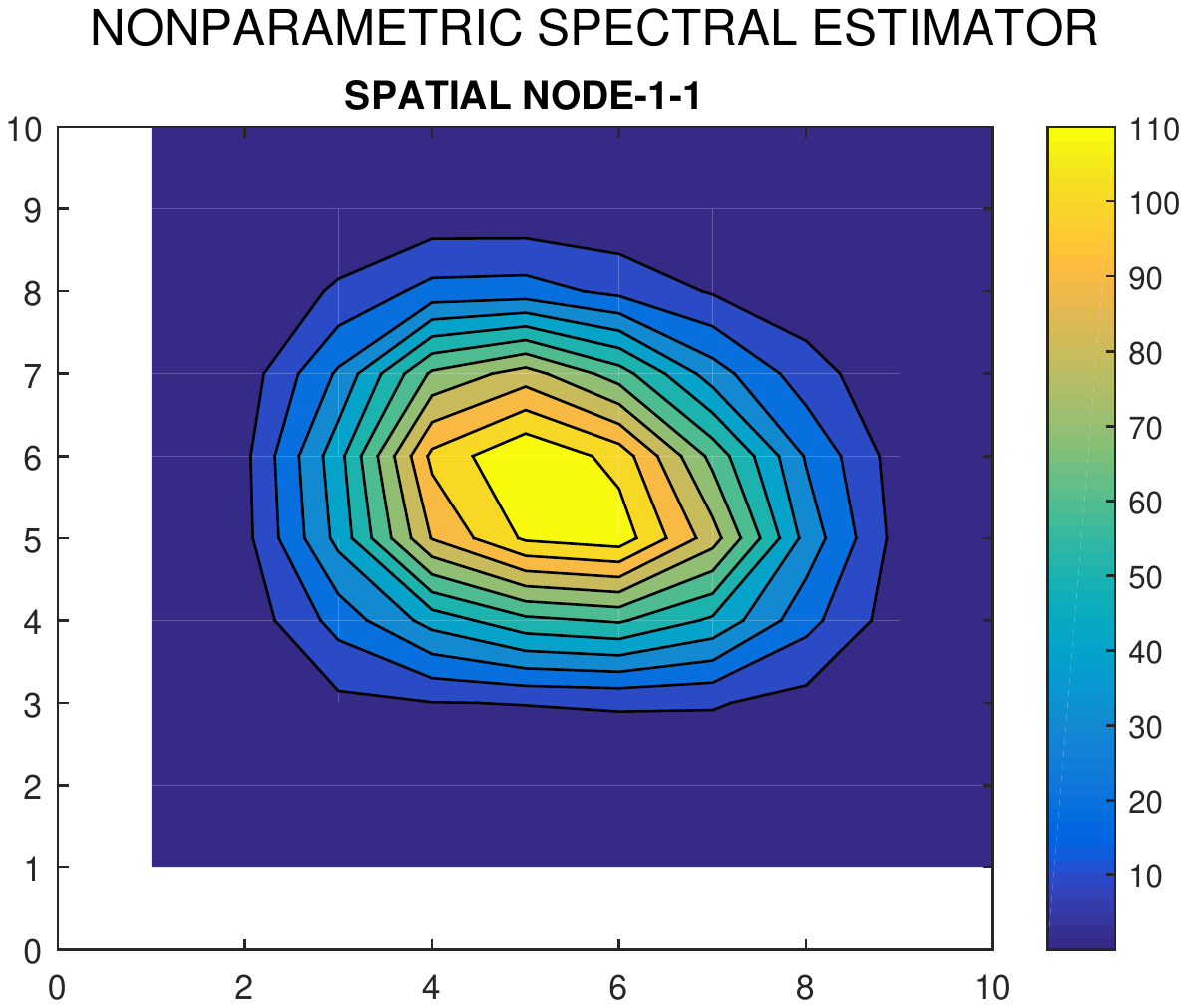}&
\hspace{-3cm}
\includegraphics[width=8.5cm,height=10.5cm]{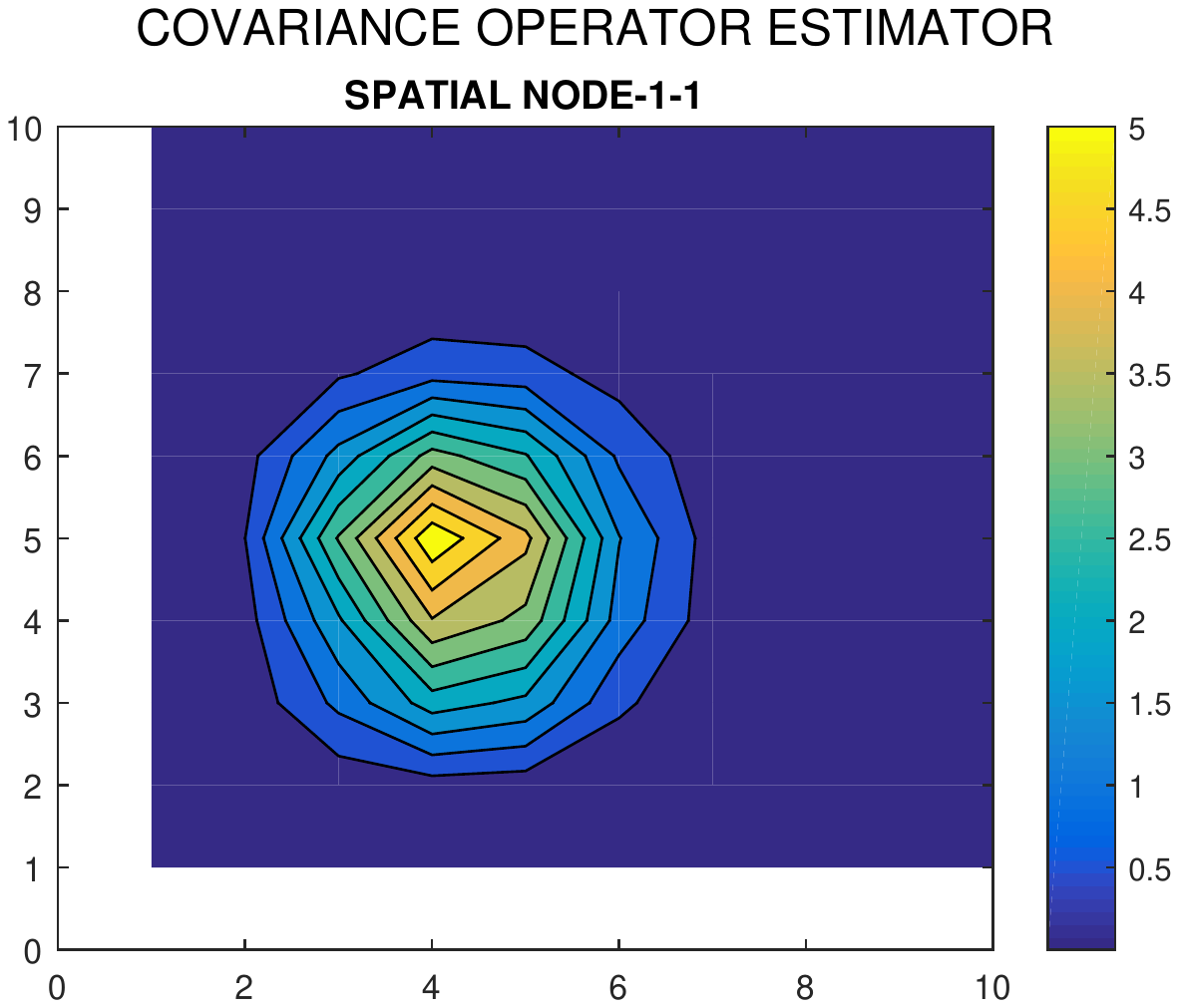}\\
\hspace{-1.5cm}
\includegraphics[width=8.5cm,height=10.5cm]{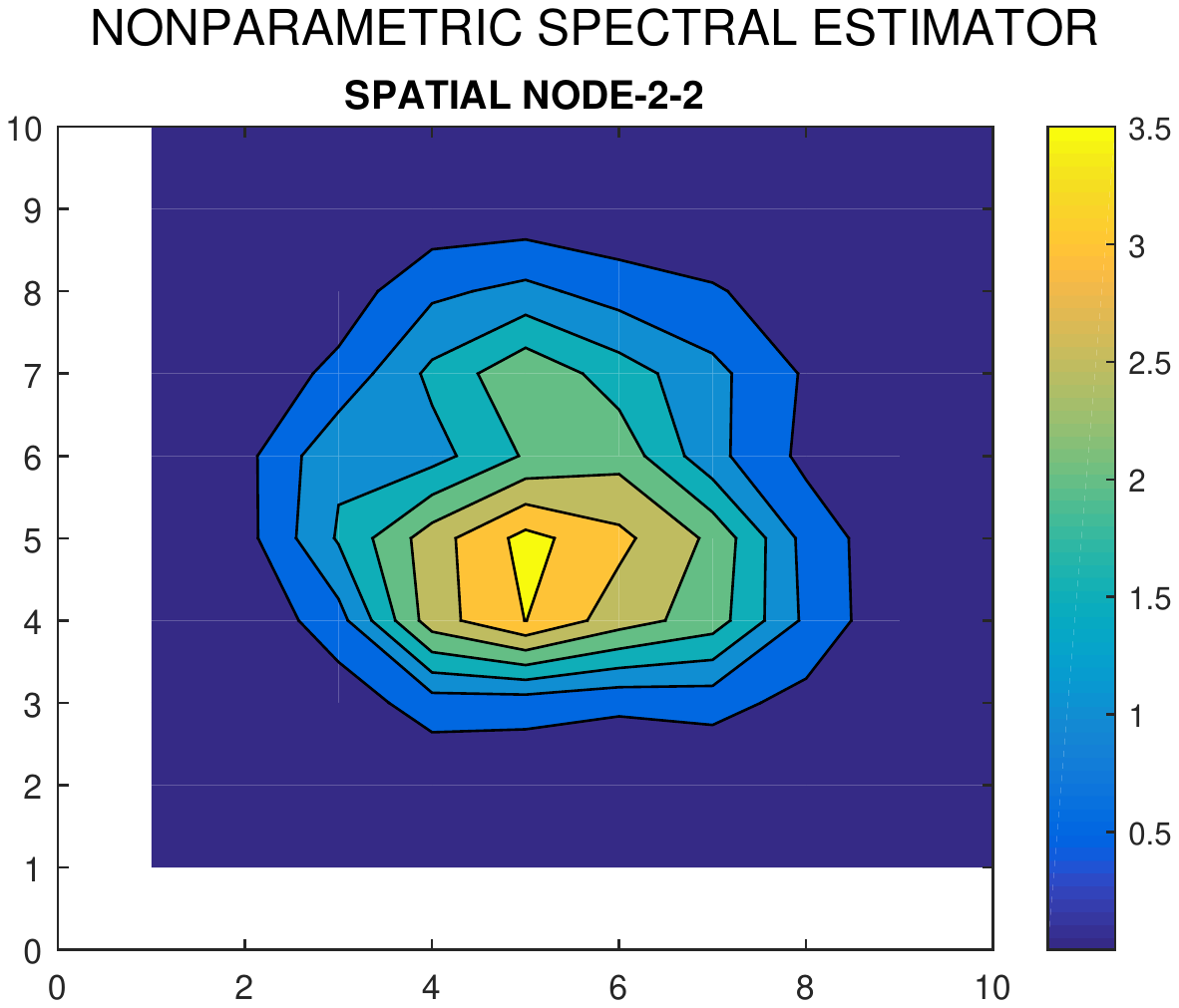}&
\hspace{-3cm}
\includegraphics[width=8.5cm,height=10.5cm]{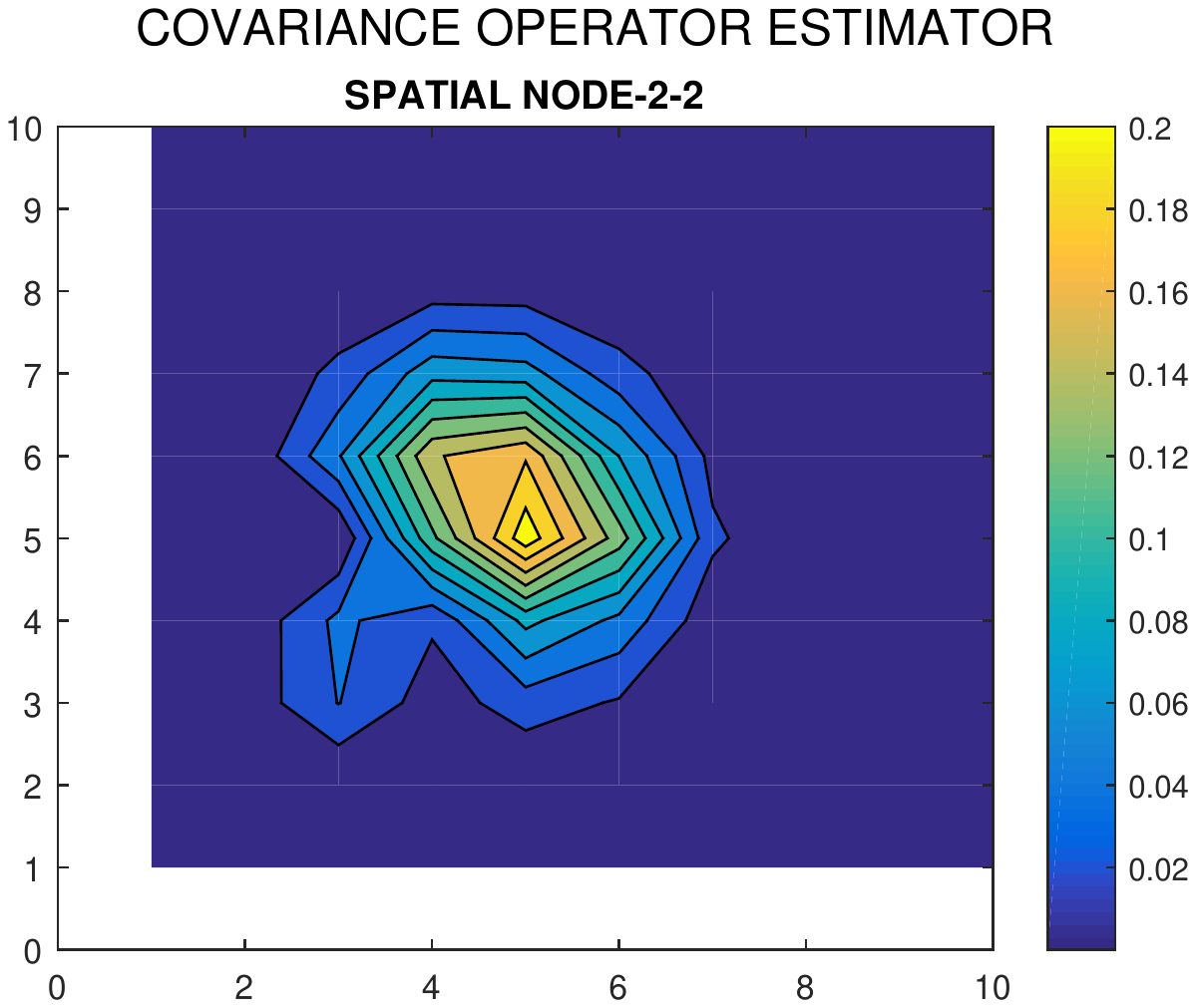}\\
\end{array}$
\end{center}\vspace*{-3cm}\caption{\scriptsize{\emph{The projected  nonparametric estimator of the spectral density operator
$\widehat{f}_{\boldsymbol{\omega}}^{(\mathbf{N})}(\psi_{k})(\psi_{l}),$ for $k=l=1$  (top--left--hand side), and for $k=l=2$  (bottom--left--hand side).
The corresponding projected spatial covariance operator estimates are displayed, for $k=l=1$  (top--right--hand side), and for $k=l=2$  (bottom--right--hand side) }}}
 \label{fig:PSOSVSFSEl2}
\end{figure}

\begin{figure}[H]
\vspace{-8.5cm}
\hspace{-9cm}
\begin{center}
$\begin{array}{c}
\hspace{-4.5cm}
\vspace{-16cm}
\includegraphics[width=22cm,height=25cm]{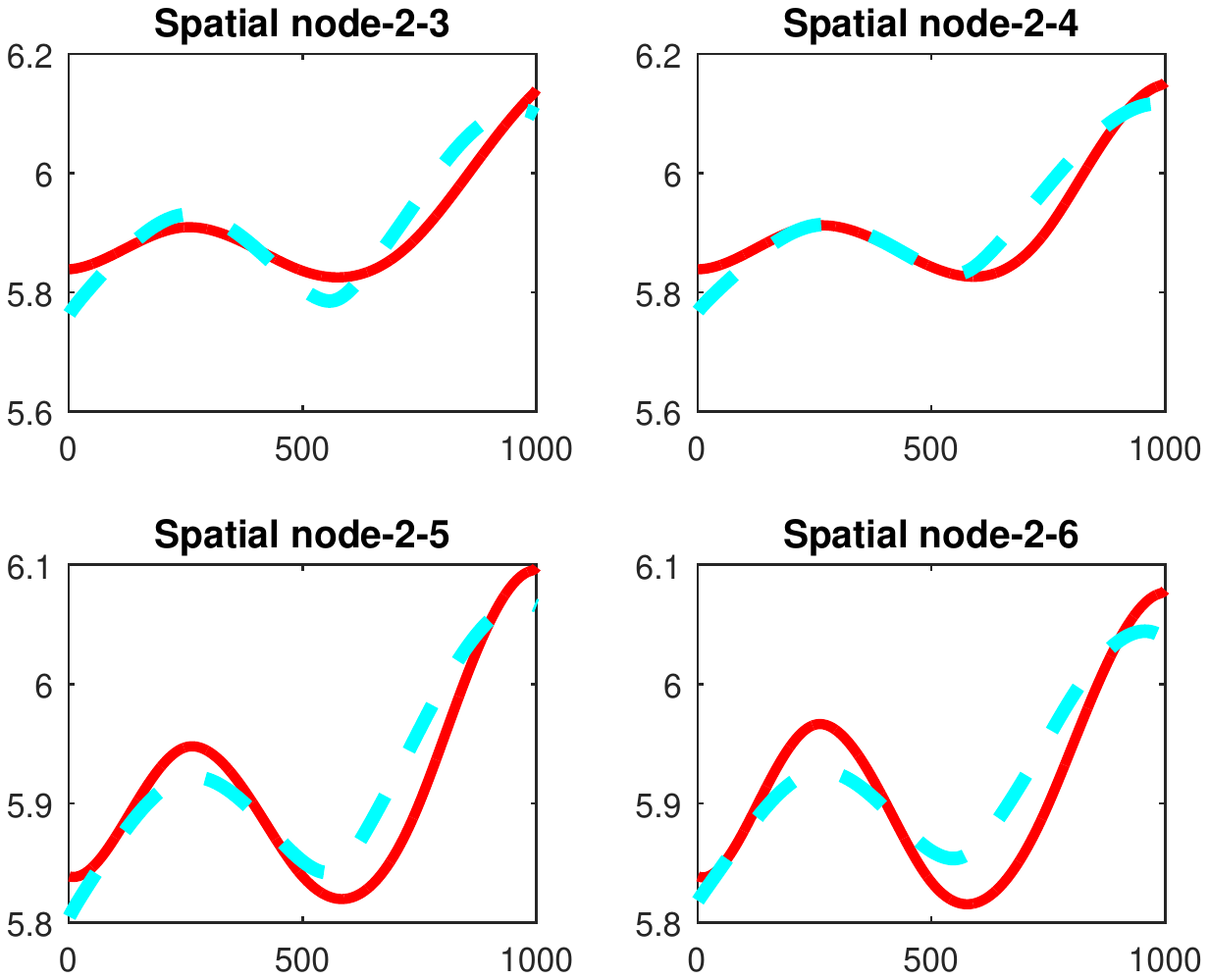}\\
\hspace{-4.5cm}
\includegraphics[width=22cm,height=25cm]{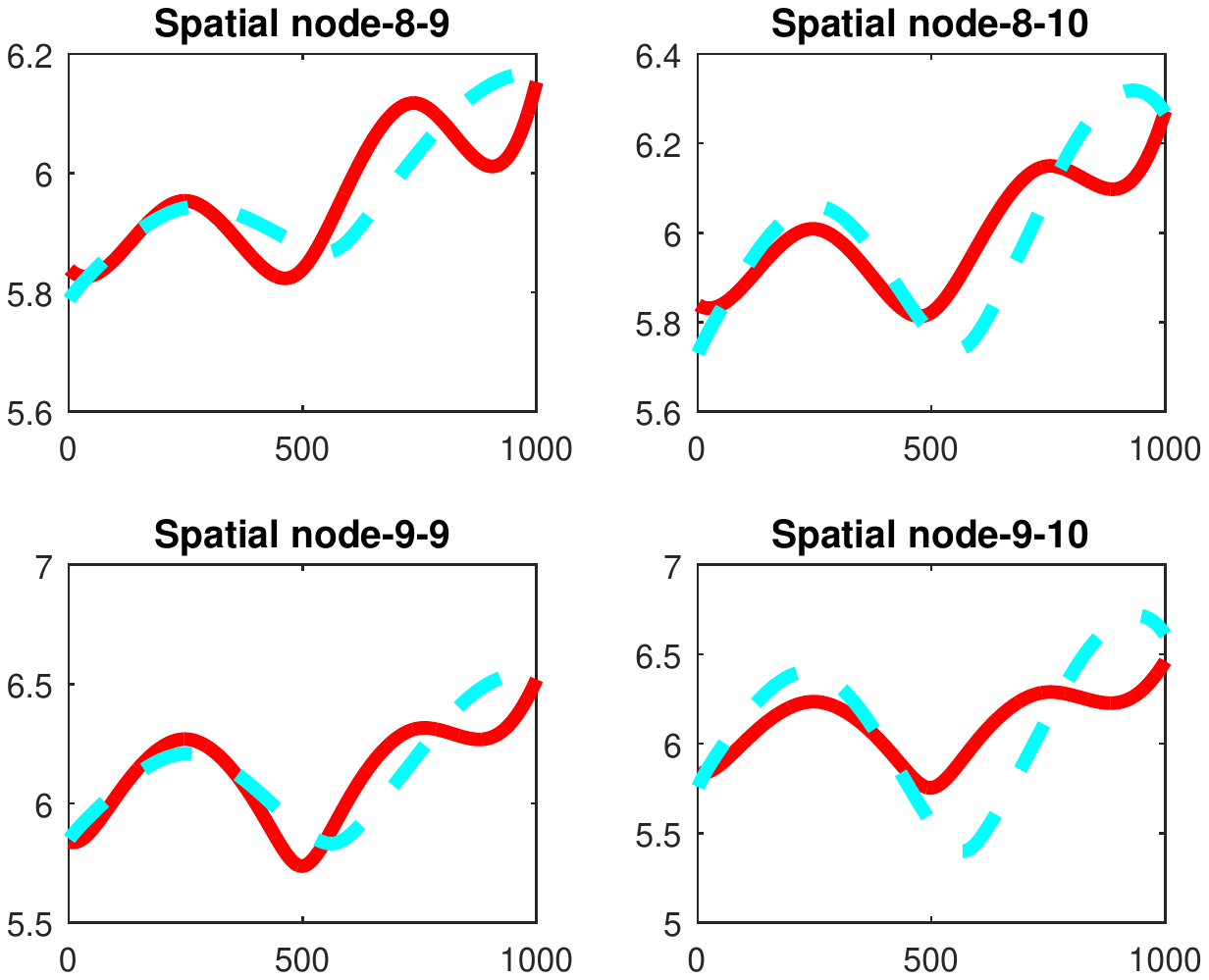}
\end{array}$
\end{center}\vspace*{-8.5cm}\caption{\scriptsize{\emph{The original curve value (red line), and its spatial functional spectral estimate (dashed blue line) are displayed at the spatial nodes $(2,3), (2,4), (2,5),$  $(2,6),$  $(8,9),$ $(8,10),$ $(9,9)$ and $(9,10).$}}}\label{fig:CESTORIGINl}
\end{figure}

As commented in Section \ref{ea1}, the data preprocessing procedure applied in  \textbf{Step 1} of the estimation algorithm 2 is almost the same to the one applied in algorithm 1, considering, in addition,    data tapering, which improves computations of  the spatial functional spectral estimators.  \textbf{Step 2} is then implemented after detrending the data. Specifically, denoting  by $X$  the  detrended data,   the  empirical long--run  spatial covariance operator $\widehat{\mathcal{R}}^{X}_{(\mathbf{N})}=\sum_{\mathbf{z}\in [0,T-1]^{d}}\widehat{\mathcal{R}}_{\mathbf{z}}$    is computed for $d=2,$   from the empirical spatial covariance operators (see also Figure 13 in the Supplementary Material):
\begin{equation}
\widehat{\mathcal{R}}_{\mathbf{z}}= \frac{1}{\prod_{i=1}^{d}T_{i}-z_{i}}\sum_{y_{i}\geq z_{i}, i=1,\dots,d}X_{\mathbf{y}}\otimes X_{\mathbf{y}-\mathbf{z}},\quad \mathbf{z}\in [0,T-1]^{d}.
\label{eqcomelrco}
\end{equation}
As output of  \textbf{Step 3}, the singular value decomposition of $\widehat{\mathcal{R}}^{X}_{(\mathbf{N})}$ is obtained by calculating
the empirical right $\{\psi_{k}^{(\mathbf{N})}\}_{k\geq 1},$ and left  $\{\varphi_{k}^{(\mathbf{N})}\}_{k\geq 1}$ eigenvectors, and the corresponding singular values $\{\lambda_{k}(\widehat{\mathcal{R}}^{X}_{(\mathbf{N})})\}_{k\geq 1}$ satisfying
$$\widehat{\mathcal{R}}_{(\mathbf{N})}^{X}\psi_{k}^{(\mathbf{N})}=\lambda_{k}(\widehat{\mathcal{R}}^{X}_{(\mathbf{N})})\varphi_{k}^{(\mathbf{N})},\quad k\geq 1.$$\noindent For $k=1,\dots, M,$ after projection onto $\psi_{k}^{(\mathbf{N})},$ we compute \textbf{Step 4} from
\begin{equation}
\widetilde{X}_{\boldsymbol{\omega }}^{(\mathbf{N})}(\psi_{k}^{(\mathbf{N})}) = ((2\pi)^{d} \mathbf{N})^{-1/2}\sum_{\mathbf{z}\in  [1,T]^{d}\cap  \mathbb{Z}^{d}} X_{\mathbf{z}}(\psi_{k}^{(\mathbf{N})})\exp\left(-i\left\langle \boldsymbol{\omega}, \mathbf{z}\right\rangle \right)
\label{dft}
\end{equation} \noindent for $\boldsymbol{\omega} \in \left\{2\pi \mathbf{z}/T,\ \mathbf{z}\in [1,T-1]^{d}\right\},$
 where the truncation parameter value $M=5$ has been selected corresponding to a $99\%$ of the empirical variability \linebreak $\sum_{k=1}^{\mathbf{N}}\lambda_{k}(\widehat{\mathcal{R}}^{X}_{(\mathbf{N})}).$
In \textbf{Step 5}, we obtain the corresponding projected periodogram operator
\begin{eqnarray}
\mathcal{I}^{(\mathbf{N})}_{\boldsymbol{\omega}}(\psi_{k}^{(\mathbf{N})})(\psi_{l}^{(\mathbf{N})})=\widetilde{X}_{\boldsymbol{\omega}}^{(\mathbf{N})}(\psi_{k}^{(\mathbf{N})})
\overline{\widetilde{X}_{\boldsymbol{\omega}}^{(\mathbf{N})}(\psi_{l}^{(\mathbf{N})})},\quad  k,l\in \{1,\dots,M\},\nonumber\\
\end{eqnarray}\noindent for $\boldsymbol{\omega} \in \left\{2\pi \mathbf{z}/T,\ \mathbf{z}\in [1,T-1]^{d}\right\}.$ In \textbf{Step 6}, the nonparametric estimator of the spectral density operator is then computed from equation (\ref{enp}), by considering a separable spatial version of the modified  Bartlett--Hann window, corresponding to  run \emph{blackmanharris} at the first argument in  the MatLab function
\emph{window}$(\cdot,\cdot )$  (see left column in Figure \ref{fig:PSOSVSFSEl2}, where two diagonal projections of the  nonparametric spectral density operator estimator are displayed, and Figure 14 in the Supplementary Material).
 \textbf{Step 7} provides the calculation of  equation (\ref{pnsdo}) in the projected spatial functional spectral domain.
\textbf{Step 8} applies \emph{ifft2}$(\cdot,\cdot)$ MatLab function to the output in \textbf{Step 7} to obtain   $\widehat{Y}_{S,\mathbf{N}}$ (see Figure
\ref{fig:CESTORIGINl}).  See also right column in  Figure \ref{fig:PSOSVSFSEl2}.
  \textbf{Step 9} is finally computed by running nine times \textbf{Steps 1--8}. Specifically, for  $n=1,\dots, 9,$ at the $n$th iteration, after removing the $n$th row and $n$th column,   \textbf{Steps 1--8}  are run from the remaining spatial curves  defining  the training sample.  After evaluate the absolute errors obtained at each iteration, by comparing the output of  \textbf{Steps 1--8}  with the  target curve sample, the mean over the nine iterations defines the curve absolute cross--validation errors over  a $9\times 9$ grid (see Figures 15, and Tables 1--4 of the Supplementary Material). Here, the average over the 1061 temporal nodes of the  pointwise values  of the absolute  cross--validation errors   are displayed in Table \ref{tabTMCVE1}. See also Figure 16 in the Supplementary Material.
   Note that the mean   of the  pointwise values of the curve absolute cross--validation errors  over the 1061 temporal nodes and 81 spatial nodes is $\mathbf{0.012789241}.$
\vspace*{-3.5cm}

\clearpage

{\scriptsize  \begin{table}[H]
\centering \resizebox{13cm}{!}{\begin{tabular} {l||c c c c c c c c c}
TIME &	C1	&	C2	&	C3	&	C4	&	C5	&	C6	&	C7	&	C8	&	C9	\\	\hline	\hline
R1 & 5.7030092e-04 & 6.8348935e-04 & 7.3573629e-04 & 1.2111422e-03 & 1.5253610e-03  & 1.1582524e-03 &  6.4996599e-04 &  4.4853841e-04  & 4.8335842e-04\\
R2 & 6.8117054e-04 & 1.0439449e-03 & 2.3038393e-03 & 4.3216034e-03 & 5.6747546e-03  & 4.0800204e-03 &  1.9086671e-03 &  9.3589234e-04 &  6.3663570e-04\\
R3 & 1.1343173e-03 & 4.9980942e-03 & 1.0964759e-02 & 2.1721765e-02 & 1.3553450e-02  &  1.2758592e-02 &  7.1904923e-03 &  2.7851634e-03 &  1.1143095e-03\\
R4 & 1.9814737e-03 & 7.4492915e-03 & 1.8550020e-02 & 5.9885317e-02 & 1.7065638e-01  & 4.3752536e-02 &  1.2645345e-02 &  5.5759098e-03 &  1.3117121e-03\\
R5 & 1.6909117e-03 & 9.1699655e-03 & 2.1533640e-02 & 3.8610072e-02 & 7.9602583e-02  & 1.4071528e-01 &  2.3334549e-02 &  7.4936816e-03 &  1.9115134e-03\\
R6 & 1.7947759e-03 & 8.3489455e-03 & 2.4934251e-02 & 3.4649351e-02 & 1.8313565e-02  &  2.7779579e-02 &  1.8405330e-02 &  6.5693512e-03 &  2.0597864e-03\\
R7 & 1.2888168e-03 & 5.1579564e-03 & 1.5046107e-02 & 2.1652733e-02 & 1.7133585e-02  &  8.0797335e-03 &  7.9961038e-03 &  3.1094061e-03 &  1.1663059e-03\\
R8 & 6.9762041e-04 & 1.7261479e-03 & 4.6013923e-03 & 9.2542787e-03 & 1.1597638e-02  & 7.8890004e-03 &  5.9333870e-03 &  2.0044843e-03 &  8.3627742e-04\\
R9 & 5.7529370e-04 & 6.4731393e-04 & 1.4868305e-03 & 3.0378125e-03 & 4.5606010e-03  & 4.1239116e-03 &  1.1999598e-03 &  6.9900174e-04 &  4.2810584e-04\\
\end{tabular}
}
\caption{\scriptsize{\emph{Average over the 1061 temporal nodes  of the absolute cross--validation errors on a $9\times 9$ grid}}}\label{tabTMCVE1}
\end{table}
}

\section{Final comments}
\label{conclusion}
This paper proposes two estimation methodologies in the context of functional regression.  The first one is  based on a bayesian approximation to the functional temporal correlation structure driving  a surface  functional time series analysis of spatiotemporal data. Here, our analysis is focused on computing the functional regression predictor of dynamical COVID--19  incidence maps at some  Spanish Autonomous  Communities. In this analysis,  LOOCV absolute errors are computed to test the suitability of the  prediction  methodology proposed in an infinite--dimensional multivariate functional regression framework.

In a second place, we adopt a spatial curve time series framework to predict COVID--19 incidence from the estimation of the spatial curve correlation structure in the spectral domain. For dimension reduction in the time domain, projection onto the empirical  long--run spatial covariance operator eigenvectors  is achieved. It can be observed that  the most significative spatial correlations through time are kept at the projections corresponding to the empirical eigenvectors associated with the largest singular values, explaining a $99\%$ of the empirical variability. Indeed, the inverse spatial functional Fourier transform of the computed nonparametric estimator of the spectral density operator keeps  the most  significative correlation values at the diagonal projections. This projected correlation structure decays for cross projections, and goes to zero relatively fast, when we consider  projections involving the empirical eigenvectors associated with the smallest empirical singular values.

To measure the predictive capability of the two functional regression  approaches cross--validation is applied. In the surface regression framework,  LOOCV is implemented by computing, according to equation  (\ref{modelreg}), for  $n=1,\dots 993,$ the surface regression  predictor $\widehat{\mathbf{Y}}_{B,N,n}$ in (\ref{predb}) at time $n,$   from the  componentwise Bayesian estimate of the temporal surface correlation structure, based on the remaining 992 surfaces defining the training sample. This predictor is evaluated and compared with the target surface  located at the temporal node $n,$ eliminated in the definition of the surface training sample, for  $n=1,\dots 993.$  Note that  $61$ surfaces from the initial sample of $1061$ surfaces  are eliminated  to remove edge effects. The remaining  $7$ surfaces at the initial times conform the random initial condition structure  required to run equation (\ref{modelreg})  with $p=7.$

A $9$--fold cross validation technique is implemented  to test the predictive performance of estimation algorithm 2.   The spatial geometric characteristics of our functional sample requires us to design a different cross--validation strategy. Specifically,  the curves at the spatial nodes of the first row and column are needed to conform our random initial condition structure. The remaining spatial curves are split into a training and validation samples at each one of the iterations of the cross--validation procedure. Thus, we compute the empirical long--run spatial covariance operator and the non--parametric estimator of the spectral density operator from the training sample, as well as the corresponding  curve regression predictor according to equation (\ref{pnsdo}) (see also Section \ref{ea2}). Here, again, this predictor is compared with the target curves located at the $n$th row and $n$th column, for each one of the  $n=1,\dots,9$ iterations.

We think that the   worst performance observed in the implementation of  the Bayesian surface regression  is  due to the more complex structure of the estimation methodology adopted, involving high--dimensional hyparameters and parameter to be fitted.   The spatial Markovian nature   of the curve data  analyzed allows us an easy and fast  implementation of the spatial functional spectral approach. Specifically,  dimension  reduction by projection of the  curve data onto the empirical eigenvectors of the long--run
spatial covariance operator, and the faster computation speed  obtained,  replacing  convolutions  by products in the spatial functional spectral domain, favors  the linear  functional regression filter calculations under this approach.

\subsection*{Acknowledgements}
This work has been supported in part by project  MCIN/ AEI/PGC2018-099549-B-I00  and CEX2020-001105-M MCIN/AEI/10.13039/501100011033.

\end{document}